2

\documentclass{emulateapj}
\usepackage{amssymb,amsmath}

\newcommand{\kms}{km\,s$^{-1}$}
\newcommand{\cosf}{$^{12}$CO(J=6--5)}
\newcommand{\cott}{$^{12}$CO(J=3--2)}
\newcommand{\coto}{$^{12}$CO(J=2--1)}
\newcommand{\cooz}{$^{12}$CO(J=1--0)}
\newcommand{\hcopft}{HCO$^+$(J=4--3)}

\slugcomment{Not to appear in Nonlearned J., 45.}

\shorttitle{Extended \cosf\, emission in NGC~253}
\shortauthors{Krips et al.}

\begin{document}


\title{SMA Observations of the Extended \cosf\, Emission in
the Starburst Galaxy NGC~253}


\author{M.~Krips\altaffilmark{1}, S.~Mart\'in\altaffilmark{1,2},
  A.~B.~Peck\altaffilmark{3}, K. Sakamoto\altaffilmark{4},
  R. Neri\altaffilmark{1}, M. Gurwell\altaffilmark{5},
  G. Petitpas\altaffilmark{5}, and Jun-Hui Zhao\altaffilmark{5}}


\altaffiltext{1}{Institut de RadioAstronomie Millim\'etrique, 300 rue
de la Piscine, Domaine Universitaire, 38406 Saint Martin d'H\`eres,
France; krips,smartin,neri@iram.fr}

\altaffiltext{2}{European Southern Observatory, Avda Alonso de
C\'ordova 3107, Vitacura, Santiago, Chile}

\altaffiltext{3}{National Radio Astronomy Observatory, 520 Edgemont
Rd, Charlottesville, VA 22903, USA; apeck@nrao.edu}

\altaffiltext{4}{Academia Sinica, Institute of Astronomy and
Astrophysics, Taiwan; ksakamoto@asiaa.sinica.edu.tw}

\altaffiltext{5}{Harvard-Smithsonian Center for Astrophysics,
Cambridge, MA 02138; mgurwell,gpetitpa,jzhao@cfa.harvard.edu}





\begin{abstract}
  
  We present observations of the \cosf\, line and 686~GHz continuum
  emission in NGC~253 with the Submillimeter Array at an angular
  resolution of $\sim$4$''$. The \cosf\, emission is clearly detected
  along the disk and follows the distribution of the lower $^{12}$CO
  line transitions with little variations of the line ratios. A
  large-velocity gradient analysis suggests a two-temperature model of
  the molecular gas in the disk, likely dominated by a combination of
  low-velocity shocks and the disk wide PDRs.  Only marginal \cosf\,
  emission is detected in the vicinity of the expanding shells at the
  eastern and western edges of the disk. While the eastern shell
  contains gas even warmer (T$_{\rm kin}>$300~K) than the hot gas
  component (T$_{\rm kin}$=300~K) of the disk, the western shell is
  surrounded by gas much cooler (T$_{\rm kin}$=60~K) than the eastern
  shell but somewhat hotter than the cold gas component of the disk
  (for similar H$_2$ and CO column densities), indicative of different
  (or differently efficient) heating mechansisms. The continuum
  emission at 686~GHz in the disk agrees well in shape and size with
  that at lower (sub-)millimeter frequencies, exhibiting a spectral
  index consistent with thermal dust emission. We find dust
  temperatures of $\sim$10-30~K and largely optically thin
  emission. However, our fits suggest a second (more optically thick)
  dust component at higher temperatures (T$_{\rm d}>60$~K), similar to
  the molecular gas. We estimate a global dust mass of
  $\sim$10$^6$~M$_\odot$ for the disk translating into a gas-to-dust
  mass ratio of a few hundred consistent with other nearby active
  galaxies.

\end{abstract}


\begin{deluxetable*}{cccccccc}
\centering
\tablewidth{0pt}
\tablecaption{Observation Log}
\tablehead{
\colhead{Line} & \colhead{Frequency} & \colhead{Date} & \colhead{rms$^a$}& \colhead{$\Delta$v}  & \colhead{Beam$^b$} & \colhead{Type$^c$}
       & \colhead{Ref$^d$} \\
               & (GHz)               &  (mm/yyyy)     &    (mJy)          & (\kms)               & ($''\times''$)    &   & }
\startdata 
\cosf          & 691.473 (USB)       &   09/2007      & 7000  & 12   & 4$\farcs$2$\times$2$\farcs$1 & M      & (1)   \\
\cott$^e$      & 345.796 (LSB)       &   09/2007      &  100  & 12   & 5$\farcs$8$\times$4$\farcs$5 & M      & (1)   \\
\cott$^e$      & 345.796 (LSB)       &   09/2004      &  130  & 12   & 3$\farcs$9$\times$1$\farcs$9 & M      & (2)   \\
\coto$^f$      & 230.538 (LSB,USB)   &   2003-2005    &   30  & 12   & 1$\farcs$7$\times$1$\farcs$5 & SF     & (2) \\
\cooz          & 115.271 (USB)       &   2011         &    4  & 12   & 3$\farcs$7$\times$2$\farcs$7 & M      & (3)  
\enddata

\tablenotetext{a}{The noise given here corresponds to the thermal
  noise, as derived from the visibility weights, i.e., system
  temperatures. The noise is given per individual mosaic field (see
  text for some more discussion).}

\tablenotetext{b}{``Original'' synthesized resolution for natural
  weighting. Please note that all maps presented in this paper were
  brought to the same resolution of 4$\farcs$2$\times$2$\farcs$1 using
  a uv-taper to change the weighting for the different baselines.}

\tablenotetext{c}{M=Mosaic; SF= Single Field}

\tablenotetext{d}{ (1) this work; (2) \citet{saka11}; (3) ALMA science
  archive; project id: 2011.0.00172.S. }

\tablenotetext{e}{These two data sets were eventually combined (see
text).}

\tablenotetext{f}{These data were taken using all available
  configurations of the SMA, with baselines ranging from 8 to 509m, as
  well observed in different sidebands.}

\label{tab1}
\end{deluxetable*}

\section{Introduction}

Submillimeter interferometric observations at frequencies above
400~GHz are technically difficult and rely on very dry atmospheric
conditions.  The reduced transmissivity of the atmosphere at submm
wavelengths necessitates observing from a vantage point at high
elevation ($\gtrsim$4000~m above sea level).  Before the advent of the
Atacama Large Millimeter Array (ALMA), the only interferometer in the
world offering the capability of observations at frequencies above
400~GHz was the Submillimeter Array (SMA) located at 4000~m altitude
on Mauna Kea in Hawaii, USA. Equipped with 8 antennas of 6~m diameter
and 690~GHz receivers, the SMA provides a window on warm and hot
molecular gas and dust through the \cosf\, line and submillimeter
continuum emission with high angular resolution. The analysis of
spatially resolved warm molecular gas provides essential information
to understand the complexity of the excitation conditions, chemistry
and dynamics of the molecular gas in various environments, including
star-forming regions, active galactic nuclei (AGN) and quiescent
regions \cite[e.g.,][]{ala11,krip11,krip08,mart06}. Moreover, given
the explosion of detections of molecular gas at high redshifts in the
past decade, mostly through high rotational levels of CO, it is
essential to have a robust understanding of the physics behind the CO
ladder as a function of the energetic environments in which they are
detected \cite[e.g.,][]{ros14,papa10,papa07,weis05,cari02}. To date,
only a handful of journal publications exist based on 690~GHz
interferometric observations, most of which study galactic sources
\cite[e.g.][]{zapa13,rolf11,mats09,naka07,qi06,beut06}.
\citet{mats09} present the first 690~GHz interferometric study of
\cosf\, emission in a nearby extragalactic source, the ultra-luminous
infrared galaxy Arp~220. Since then, mostly thanks to ALMA but also
the SMA, a couple of interferometric observations of the \cosf\,
emission in nearby galaxies have been very recently published
\cite[e.g.,][]{garc14,xu14,sliw13}. In this paper, we present the
first mosaic observations at 690~GHz of the nearby starburst galaxy
NGC~253, made using the SMA.

NGC~253 is one of the best studied nearby \cite[$\sim$3.5 Mpc, where
1$''$=17~pc;][]{reko05,mouh05} infrared bright starburst galaxies
\cite[3$\times$10$^{10}$L$_\odot$;][]{tele80}. It is thus an ideal
prototype to study the effects of a central starburst
\cite[$\sim$2-4M$_\odot$~yr$^{-1}$;][]{minh07,ott05,bendo2015} on the
dynamics (NGC~253 is a edge-on galaxy), excitation conditions and
chemistry of the surrounding molecular gas
\cite[e.g.,][]{leroy15,meier15,bola13,saka11,saka06,knud07,mart06,brad03}.
As a consequence of the starburst in the center of NGC~253, a high
rate of supernova explosions ($<$0.2~yr$^{-1}$) has been observed
\cite[see][]{rampa14,pagl12}.  These are most likely responsible for a
kiloparsec-scale outflow perpendicular to the disk
\cite[e.g.,][]{fabb84,heck90}. In addition to this outflow,
\citet{saka06} report on at least two expanding shells, or
superbubbles, at the edges of the disk (named SB1 for the
south-western shell and SB2 for the north-eastern one) that might be
either caused by winds from and/or supernovae in a super star cluster
or a hypernova(e); the possibility of a third shell, a bit more
southern to SB2, was discussed as well. The authors further find a
strongly disturbed molecular gas disk associated with young stellar
clusters and stellar explosions as well as the large-scale
super-wind. Their follow-up observations at $\sim$1$"$ resolutions
confirmed the SB2 (and the third shell) but SB1 was reported to be
more complicated than appeared in previous lower-resolution data
\cite[][]{saka11}. They noted that it may be a shell in a complicated
shape or other gas kinematical feature that looks like a shell at low
resolution.  With very recent ALMA observations of \cooz\, at 3$"$
resolution \cite{bola13} reported four shells in the central kpc of
NGC~253. One is SB2 and three are clustered in/around the region of
SB1. While SB2 can be connected to a compact stellar cluster, SB1 is
found at the position of a compact radio continuum source that is
associated with a supernova remnant \cite[see][]{saka06,bola13}. The
nature of shells 2 and 4 remain unclear at this point, whether they
are separate bubbles or connected to the stellar winds or other
dynamical features.

A 2~mm line survey of NGC~253 done by \citet{mart06} reveals an
impressive chemical richness and complexity in its center \cite[see
also][]{meier15}. NGC253 appears to resemble SgrB2, the region of
molecular cloud complexes in the Galactic center, more than the
evolved starburst galaxy M82 \citep[see also][]{ala11} despite the
fact that our Milky Way is not a starburst galaxy at all. Moreover,
\citet{saka11} find a striking resemblance between the molecular gas
distribution of the Milky Way and that of NGC~253. The chemistry of
the molecular gas in the central kpc of NGC~253 appears to be
dominated by large-scale, low velocity shocks
\citep[e.g.,][]{mart06,garc00}. However, \cite{mart09} detect
significant amounts of molecular tracers for photodissociation regions
(PDRs) suggesting PDR chemistry also plays a significant role in this
starburst galaxy.  This is further supported by \citet{ros12} and
\cite{ros14}. The NIR H$_2$ emission appears to be mostly
fluorescently excited favoring PDRs as the dominant excitation
mechanism in the nuclear region of NGC~253. The only exceptions seem
to be three small isolated regions in which shocks may play the
leading role. \cite{ros12} estimate that at most 30\% of the H$_2$
emission is excited by shocks. Based on a $^{12}$CO and $^{13}$CO
ladder analysis from Herschel and ground-based single-dish
observations with a (normalised) beam of 32.5$"$, \cite{ros14} present
even more evidence for the combined contribution of PDR and mechanical
heating of the molecular gas in the central disk of NGC~253. They also
argue that heating by cosmic rays can be mostly neglected; at most a
few percent could be attributed to cosmic ray heating in their models.

The paper is organised as follows: The observations and the archival
data are discussed in Section~\ref{obs}. The results are discussed in
Section~\ref{res} and a summary is given in Section~\ref{sum}.

\section{Observations}
\label{obs}

First, we will present some general information on the SMA
observations conducted for this paper. We will then briefly discuss
previous \cott\, SMA observations that were eventually used to merge
with the \cott\, observations carried out simultaneously with
\cosf. We will give additional information on previous \coto\, SMA
observations. These observations were taken from \cite{saka11} and
\cite{saka06}. New \cooz\, ALMA observations of NGC~253 are also
included in this study and are hence briefly presented as well. We
will address the problem of missing short spacings and the
significance of spatial filtering to our observations in
Section~\ref{shortspace}.  All data presented in this paper stem from
mosaic observations (except the \coto\ data) and have hence been
corrected for the respective primary beams.

\begin{deluxetable}{crrrr}
\centering
\tablewidth{0pt}
\tablecaption{General Results}
\tablehead{
\colhead{Line} & \colhead{$S_\nu^{\rm max,a}$} & \colhead{$v_0$$^{\rm max,a}$} & \colhead{FWHM$^{\rm max,a}$} & \colhead{$\int\left(S_\nu~dv\right)$$^{\rm b}$} \\
    $^{12}$CO   & (Jy)              &  (\kms)         & (\kms)         & (Jy~\kms) }
\startdata 
J=6--5      & 350$\pm$15            & 0$\pm$3      & 130$\pm$7      & (58$\pm$3)$\times$10$^3$      \\
J=3--2      & 347$\pm$3             & 0$\pm$1      & 140$\pm$7      & (75$\pm$2)$\times$10$^3$      \\
J=2--1      & 111$\pm$2             & 0$\pm$1      & 167$\pm$8      & (29$\pm$0.3)$\times$10$^3$    \\
J=1--0      &  30$\pm$1             & 0$\pm$1      & 165$\pm$8      & (7.8$\pm$0.1)$\times$10$^3$    \\
\hline
\colhead{Cont.} & \colhead{$S_\nu$$^{\rm b}$}          \\
     $\nu_{\rm obs}$          & (Jy)                  \\
\hline
686~GHz    & 32$\pm$1.4     \\
350~GHz    & 2.6$\pm$0.1   \\
220~GHz    & 0.67$\pm$0.05 \\
115~GHz    & 0.26$\pm$0.01 
\enddata

\tablenotetext{a}{Peak fluxes, zero velocities and FWHM are given for
  the dominant maximum component of the multiple Gaussian fit. The
  zero velocity is with respect to the redshifted frequency of the
  respective CO line. The redshift of NGC~253 is z=0.000811
  \cite[see][]{koribalski2004}}

\tablenotetext{b}{Emission taken over the entire emission area covered
  with the 5-point mosaic by using a multiple Gaussian fit (up to 3
  Gaussian lines are fitted).}

\label{tab2}
\end{deluxetable}

\subsection{SMA observations of \cosf\, and \cott}
NGC~253 was observed using two receivers simultaneously to obtain
\cosf, \cott\, and \hcopft\footnote{Further analysis of the \hcopft\,
  data may be presented separately, while this paper is concerned only
  with the $^{12}$CO transitions}\, with the SMA.  At the time of the
observations, five of the eight SMA antennas were equipped with
working 690~GHz receivers. All five antennas were placed in the inner
ring of the array, known as the sub-compact configuration, with
baselines ranging from a few ($\sim$8~meters) up to 25~meters to allow
for a good {\it(u,v)} coverage and reasonable angular resolution at
690~GHz. To cover most of the CO emission in NGC~253, a five-field
mosaic (size$\simeq$50$''$) along the plane of the disk was observed
with a spacing of half the primary beam size ($\sim$7.5$''$) at
690~GHz between the pointing centers. The phase reference center of
the central pointing was set to $\alpha(J2000)$=00$^{\rm h}$47$^{\rm
  m}$33$\fs$241 and
$\delta(J2000)$=$-$25$^\circ$17$'$18$\farcs$16. Considering the small
primary beam of $\sim$15$''$ at 690~GHz and thus the increased need
for accurate pointing, regular pointing offset updates (i.e., every
$\sim$3~h) were conducted on either 3C454.3 or 3C111 (using
interferometry) or Jupiter (using single-dish measurements) throughout
the night at 345~GHz.  All spectral windows covered 2GHz in bandwidth,
and the lower sideband (LSB) of the lower frequency was tuned to the
\cott\, line so that the \hcopft\, line fell into the corresponding
upper side band (USB), situated 10~GHz above the LSB. The USB of the
690~GHz receivers were tuned to the \cosf\, line leaving the
corresponding LSB for continuum measurements. The weather was
excellent with a 225~GHz atmospheric opacity of $\sim$0.05-0.1
throughout the track. This corresponds to single sideband (SSB) system
temperatures of T$_{\rm sys}$(SSB)=5000-10000~K at 690~GHz and T$_{\rm
  sys}$(SSB)=400-800~K at 345~GHz.  We used the nearby
($\sim$20$^\circ$ from NGC~253) planet Uranus as bandpass, gain and
absolute flux calibrator. The calibration on Uranus was performed
using a disk model to compensate for the fact that it was slightly
resolved. The quality of the gain calibration was verified against a
nearby quasar (J2348-1631) which was observed with the same cycle
times as Uranus. It appears to be of symmetric Gaussian shape at both
frequencies but slightly shifted ($\sim+$0.8$''$) in Declination from
its phase center at 690~GHz while being centered correctly at 345~GHz
after applying the gain calibration. The spatial shift at 690~GHz
indicates a problem with the phase transfer at 690~GHz, probably due
to uncertainties in the measured baseline and/or larger distance of
NGC~253 to the phase calibrator; the baseline error is around
0.3$\lambda$ at 690~GHz. We hence consider this effect as a systematic
instrumental and calibrational artifact and correct the data for this
$\sim$0.8$''$ shift. A similar positional shift was found for the
690~GHz SMA observations of Arp~220 that was also attributed to
baseline errors by \citet{mats09}.  The accuracy of the flux
calibration is estimated to be within 30\% at 690~GHz and $\sim$20\%
at 345~GHz.

Angular resolutions of 4$\farcs$2$\times$2$\farcs$1 at position angle
PA=162$^\circ$ are obtained at 690~GHz and of
5$\farcs$8$\times$4$\farcs$2 at PA=173$^\circ$ at 345~GHz.  We reach
an rms noise level of 1$\sigma$=100~mJy~beam$^{-1}$~channel$^{-1}$ at
345~GHz and of 1$\sigma$=7~Jy~beam$^{-1}$~channel$^{-1}$ at 691~GHz
per individual mosaic field and for a spectral resolution of
12~\kms. Note that we used a Nyquist sampling to place the mosaic
fields based on the 690~GHz field of view so that in the combined
mosaic maps at 690~GHz the noise distribution is not homogeneous and
will increase toward the edges and be a factor of $\sqrt{2}$ smaller
in the overlap regions; the mosaic fields at 345~GHz will, however,
overlap by a much larger fraction so that the noise distribution is
much more homogeneous there. The latter fact has been taken into
account in our estimate of the 1$\sigma$ levels for each figure (see
their captions).  A log of the observations is given in
Table~\ref{tab1}.

We subtracted the strong continuum emission found in NGC253, averaged
over all line-free channels in each sideband, from the original data
cubes to generate pure line emission data.

\begin{figure*}[!t]
   \centering
   \resizebox{16cm}{!}{\includegraphics{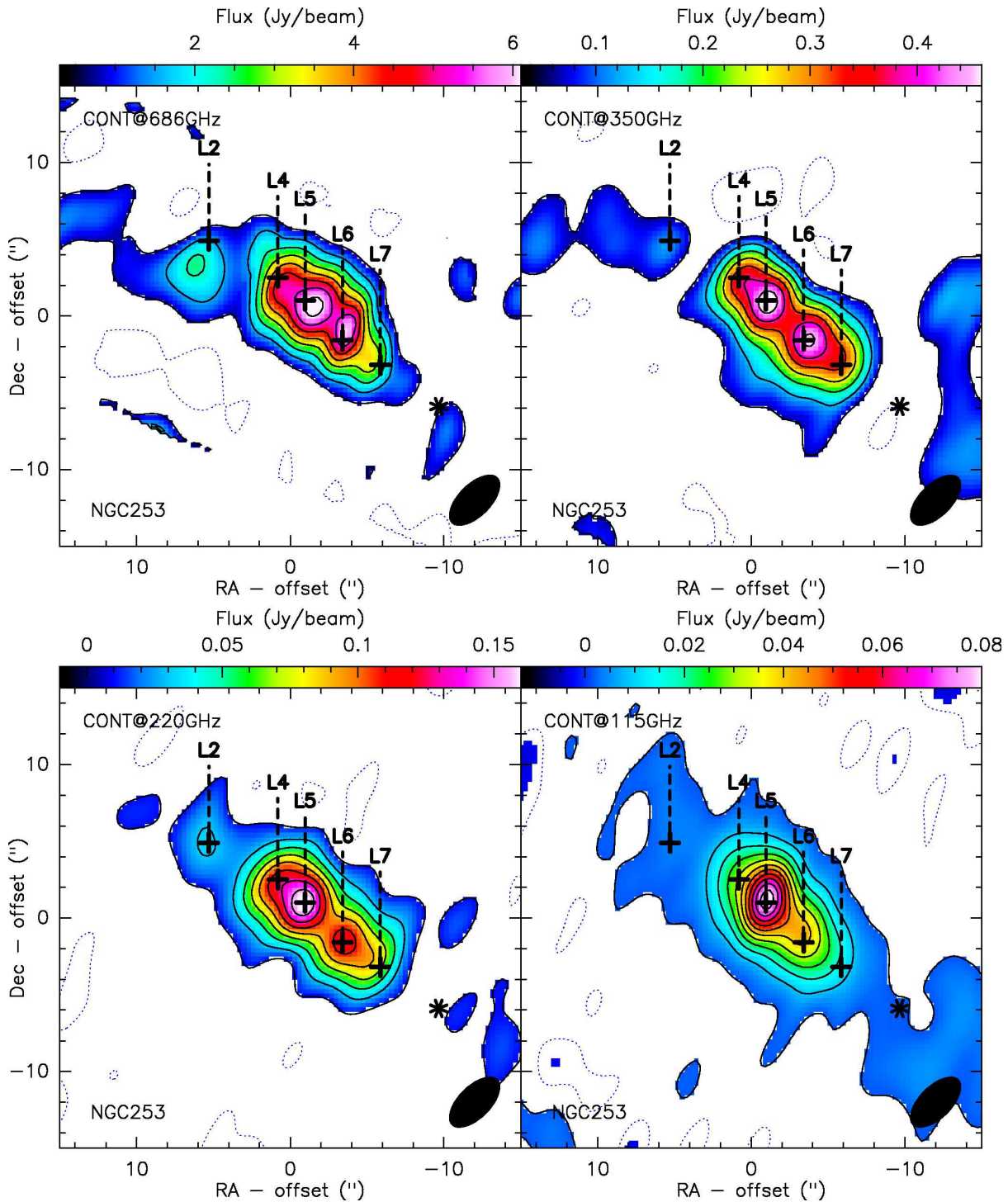}}
   \caption{Continuum emission at 686~GHz (panel 1, {\it upper left}),
     350~GHz (panel 2, {\it upper right}), 220~GHz (panel 3, {\it
       lower left}) and 115~GHz (panel 4, {\it lower right}), all at a
     spatial resolution of $\sim$4.2$''\times$2.1$''$. The contours of
     the 686~GHz continuum emission start at $-$2$\sigma$,
     2$\sigma$=0.76~Jy~beam$^{-1}$ in steps of 2$\sigma$, those of the
     350~GHz continuum emission start at $-$5$\sigma$,
     10$\sigma$=0.064~Jy~beam$^{-1}$ and go then in steps of
     10$\sigma$, those of the 220~GHz continuum emission start at
     $-$3$\sigma$, 5$\sigma$=0.01~Jy~beam$^{-1}$ and go in steps of
     10$\sigma$, and those of the 115~GHz continuum emission start at
     $-$10$\sigma$, 10~$\sigma$=1.5~mJy~beam$^{-1}$ and go in steps of
     50$\sigma$. The black crosses indicated in each panel mark the
     positions of the respective line emission peaks defined in
     Fig.~\ref{fig3} matching those of the continuum peaks shown here
     and being in good agreement with the 1.3~mm continuum emission
     peaks from \citet{saka11}. The black star marks the position of
     the western superbubble/shell from \citet{saka06}. Note, that the
     eastern superbubble/shell lies outside the maps and is not shown
     here since no continuum emission is detected around it.  The
     (0$"$,0$"$) offset is relative to the following absolute
     position: $\alpha(J2000)$=00$^{\rm h}$47$^{\rm m}$33$\fs$241 and
     $\delta(J2000)$=$-$25$^\circ$17$'$18$\farcs$16.}
   \label{fig1}
\end{figure*}

\subsection{Previous SMA observations of \cott\, and \coto}
In order to enable a direct comparison between the \cosf\, emission
and the lower-J CO emission and reduce systematic biases such as
resolution effects due to different synthesized beams and
uv-coverages, we merged our \cott\, data with the high spectral
resolution data at 345~GHz presented by \citet{saka11}. In the merged
\cott\, data set we then reach an rms noise level of
1$\sigma$=80~mJy~beam$^{-1}$~channel$^{-1}$ per mosaic field or
$\sim$60~mJy~beam$^{-1}$~channel$^{-1}$ in the center of the mosaic
for a spectral resolution of 12~\kms.  We also included the \coto\,
data by \citet{saka11} and \cite{saka06} which is already a
combination of low and high spatial resolution observations; the rms
is found to be 25~mJy~beam$^{-1}$~channel$^{-1}$ (using only a single
field) at a spectral resolution of 12~\kms. In order to match the
resolution of \cott\, and \coto\, to that of \cosf, we used a
uv-taper, obtaining an angular resolution of
4$\farcs$2$\times$2$\farcs$1. We did not find significant differences
with this method compared to matching\footnote{i.e., restricting the
  uv-coverages to the overlapping regions} the uv-coverages of the
\cott\, and \coto\, data to that of \cosf. In order to preserve the
data in its entirety, we hence chose the first method.

\subsection{ALMA observations of \cooz}

ALMA cycle~0 observations are available for the \cooz\, line from the
ALMA science archive, that were done in the compact configuration with
up to 19 antennas (project id: 2011.0.00172.S; see also
\cite{bola13}). The \cooz\, line was tuned to the USB within a 2~GHz
large spectral window using a spectral resolution of 488~kHz
($\simeq$1.3~\kms). Uranus was used as absolute flux calibrator,
J2333-237 as a bandpass calibrator and J0137-245 as phase and
amplitude calibrator.  Using the strong continuum emission,
self-calibration was applied to the phases to improve the image
quality of the continuum and line maps. Given the excellent
uv-coverage, we used a slight uv-taper to smooth the initial spatial
resolution to the angular resolution of our \cosf\, observations. Data
reduction was done using CASA but further image processing was done in
GILDAS. As with the other datasets, the continuum was derived from the
line free channels and then subsequently subtracted from the channels
containing both line and continuum emission.

\begin{figure*}[!t]
   \centering
   \resizebox{\hsize}{!}{\includegraphics{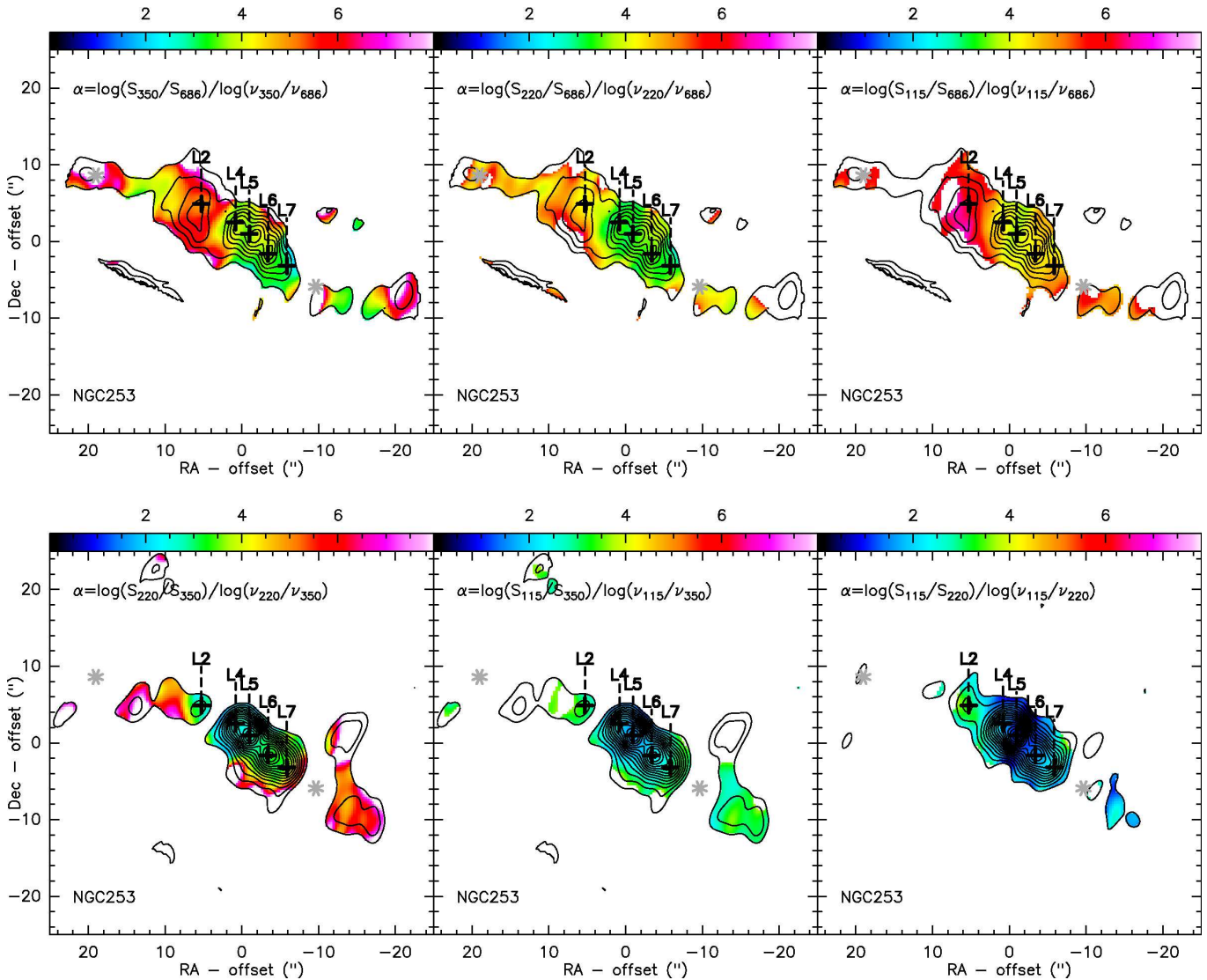}}
   \caption{Spectral emissitivity indices (labelled $\alpha$ in the
     images) derived from the continuum emission at 686~GHz, 350~GHz,
     220~GHz and 115~GHz. The contours are the same as in
     Fig.~\ref{fig1} and are from the continuum emission at the
     frequency in the denominator, respectively. The grey stars and
     black crosses are the same as in Fig.~\ref{fig1}.}
   \label{fig2}
\end{figure*}

\subsection{Effects of Spatial Filtering}
\label{shortspace}
In order to assess the lack of short spacings and hence missing flux,
we have compared our observations with existing single dish
observations. \citet{baye04} have observed the \cosf\, emission in
NGC~253 using the Caltech Submillimeter Observatory (CSO).  In the
central 10.5$''$, they obtain an integrated intensity of
1394$\pm$279\,K\,\kms, which corresponds to
$\sim$(55000$\pm$11000)\,Jy\,\kms\, assuming a conversion factor of
S/T$_{\rm mb}=$39\,Jy/K. The SMA observations yield an integrated line
intensity of $\sim$45000\,Jy\,\kms\, in the central 10.5$''$, a bit
less than that from the single-dish observations but still within the
large uncertainties ($\sim$20-30\%) in both measurements. For the
\cott\, line emission, we find integrated line intensities of
$\sim$45000\,Jy\,\kms\, in the central $\sim$20$''$ very similar to
those obtained from the single-dish measurements in the same area
discussed in \citet{baye04}. Hence, the SMA observations did not
resolve out significant emission from the \cott\, and \cosf\, lines
(at least within the central $\sim$10-20$''$)\footnote{Note that no
  single-dish observations exist to cover the entire area of the
  5-point mosaic observations conducted for this paper. The estimates
  of resolution effects are hence only for the central disk not
  covering the two outer shells}.

The \cooz\, was observed by \cite{hough97} using the SEST telescope.
They derive an integrated intensity of 377~K~\kms\, in the central
$\sim$30$''$ of NGC~253. Using a conversion factor of S/T$_{\rm
mb}$=19~Jy/K, this translates into 7100~Jy~\kms. We find
$\sim$7000~Jy~\kms\, in the central $\sim$30$''$ as well, indicating
that no flux has been resolved out with ALMA, at least in the center
of NGC~253.

\begin{figure}[!t]
   \centering
   \resizebox{7.89cm}{!}{\includegraphics{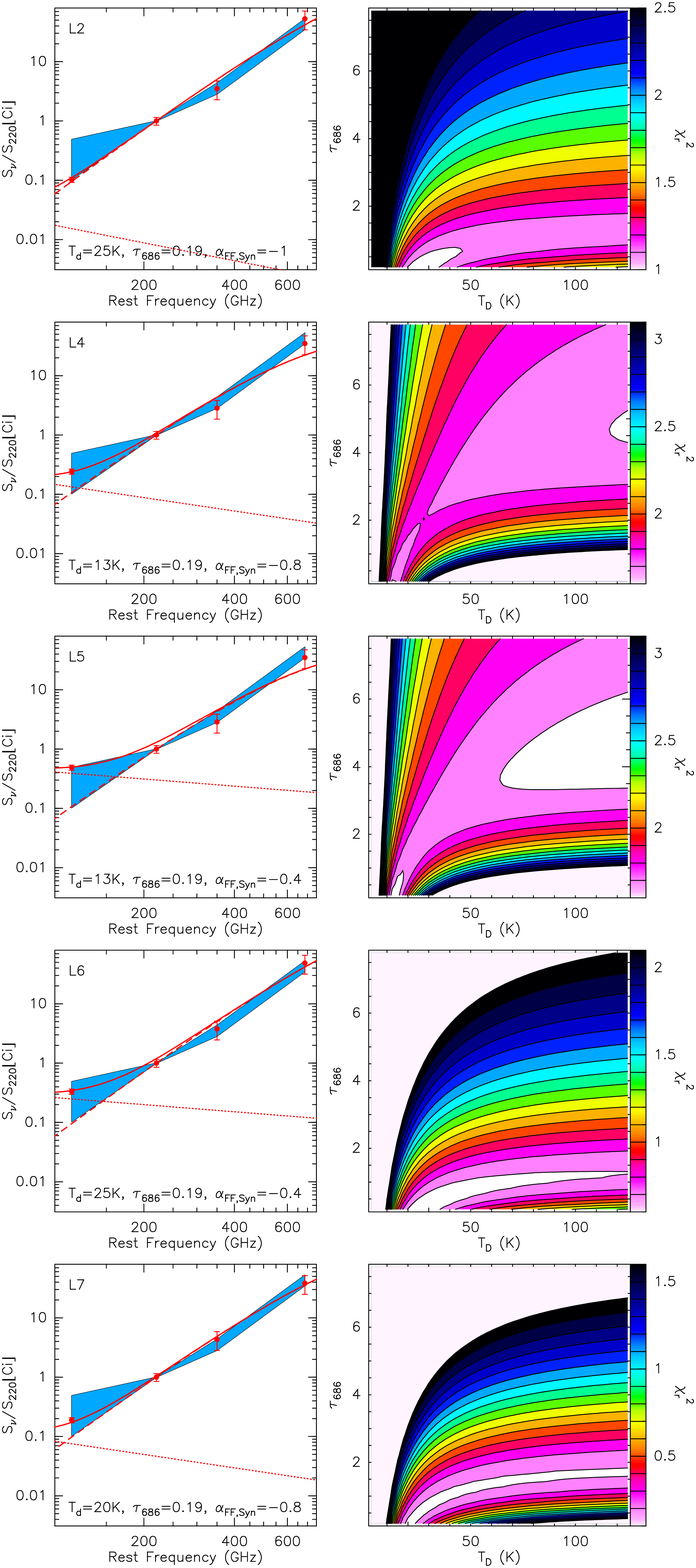}}
   \caption{Continuum SED for the different continuum peaks (only peak
     fluxes taken for each frequency; {\it left panel}), normalised to
     the respective continuum fluxes at 230~GHz for each component
     (note the logarithmic scale of the flux ratio) and the
     corresponding best fit solution (i.e., lowest reduced $\chi_{\rm
       r}^2$; {\it right panel}) for each continuum peak assuming a
     dust emissivity of $\beta=$1.8 (see text for more details). The
     dashed and dotted lines represent the thermal (from a modified
     black body) and non-thermal synchrotron plus thermal free-free
     contribution to the emission, while the solid line is the
     composite of both. The dashed line has been fitted to the data,
     while the dotted line was made using a fixed beta of 1.8 from
     \cite{hees11} (there are not enough observational data to conduct
     a thorough fit for beta as well). The error bars correspond to
     $\pm$1$\sigma$. The blue shaded areas plotted in each panel on
     the left indicate the range of continuum fluxes found at the
     positions of L2-L7, i.e., correspond to min(flux@(L2-L7)) to
     max(flux@(L2-L7)) at each wavelength. T$_{\rm d}$ is the dust
     temperature , $\tau_{686}$ is the opacity at 686~GHz and
     $\alpha_{\rm FF,Syn}$ is the spectral index coming from the
     free-free/synchrotron emission.}
   \label{fig2.5}
\end{figure}

\section{Results \& Discussion}
\label{res}

\subsection{Continuum Emission}
\label{seccont}

\begin{deluxetable}{lrrrr}[!t]
\centering
\tablewidth{0pt}
\tablecaption{Peak fluxes for the individual continuum peaks}
\tablehead{
\colhead{ID} & \colhead{686~GHz$^a$} & \colhead{350~GHz$^a$} & \colhead{220~GHz$^a$}  & \colhead{115~GHz$^a$}  \\
              & (mJy~b$^{-1}$)   & (mJy~b$^{-1}$)   & (mJy~b$^{-1}$)    & (mJy~b$^{-1}$)  }
\startdata 							  				
L2            &  1700$\pm$400  &   114$\pm$6 &    33$\pm$2  &    3$\pm$0.15 \\
L4            &   4000$\pm$400  &   326$\pm$6 &   115$\pm$2  &   28$\pm$0.15 \\
L5            &   5800$\pm$400  &   474$\pm$6 &   166$\pm$2  &   81$\pm$0.15 \\
L6            &   5900$\pm$400  &   456$\pm$6 &   121$\pm$2  &   40$\pm$0.15 \\
L7            &   2900$\pm$400  &   330$\pm$6 &   076$\pm$2  &   14$\pm$0.15 
\enddata 
 
\tablenotetext{a}{Errors are based on noise levels; b=beam (with a
  synthesizes matched beam of 4$\farcs$2$\times$2$\farcs$1 for all
  frequencies.}
\label{tab2.5}
\end{deluxetable}

\begin{figure*}[!t]
   \centering
   \resizebox{16cm}{!}{\includegraphics{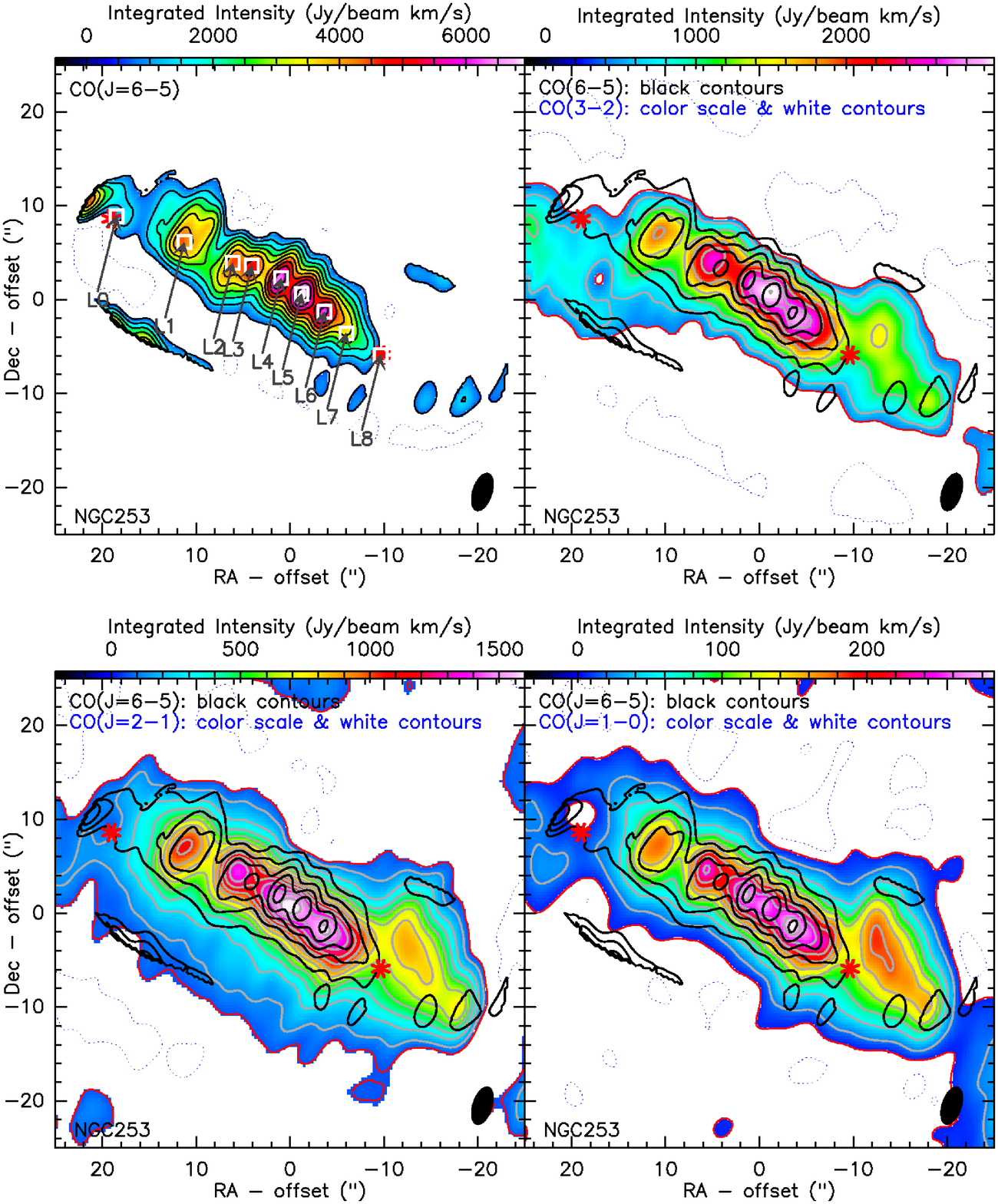}}
   \caption{Panels comparing the flux distributions of the four
     $^{12}$CO line transitions.  Panel 1 ({\it upper left}) shows the
     \cosf\, emission in color scale and contours.  Panel 2 ({\it upper
       right}) shows the \cott\, emission (this paper and
     \cite{saka11}; in {\it color scale} and {\it red and grey
       contours}) overlaid with contours (in {\it black}) of the
     \cosf\, line emission.  Panel 3 ({\it lower left}) shows the same
     but for the \coto\, emission \cite[taken from][]{saka11} overlaid
     with the contours of the \cosf\, emission. Panel 4 ({\it lower
       right}) shows the same but for the \cooz\, emission (taken from
     the ALMA science archive) overlaid with the contours of the
     \cosf\, emission. The red stars (inner and outer two white
     squares) mark the positions of the two shells (L0 and L8) from
     \citet{saka06} and L1-L7 ({\it white squares}) with the
     corresponding arrows mark the line peaks of the \cosf\, emission
     in order to facilitate the discussion. The contours of the
     \cosf\, emission start at $-$3$\sigma$,
     2$\sigma$=660~Jy~\kms~beam$^{-1}$ in steps of 2$\sigma$ (upper
     left panel, otherwise in steps of 4$\sigma$).  The contours of
     the \cott\, emission start at $-$20$\sigma$,
     100$\sigma$=420~Jy~\kms~beam$^{-1}$ in steps of 100$\sigma$. The
     contours of the \coto\, emission start at $-$50$\sigma$,
     50$\sigma$=64~Jy~\kms~beam$^{-1}$ in steps of 100$\sigma$. The
     contours of the \cooz\, emission start at $-$25$\sigma$,
     25$\sigma$=5~Jy~\kms~beam$^{-1}$ in steps of 150$\sigma$.}
   \label{fig3}
\end{figure*}

In our new SMA observations, continuum emission has been detected at
345~GHz (LSB), 355~GHz (USB), 681~GHz (LSB) and 691~GHz (USB) and is
consistent between the two sidebands acounting for the frequency
difference of 10~GHz between the sidebands. The upper left panel of
Fig.~\ref{fig1} shows the sideband-averaged continuum emission at
686~GHz, the upper right panel shows 350~GHz (merged with the data
from \cite{saka11}), the lower left panel that of 220~GHz and the
lower right panel that of 115~GHz. As mentioned before, all four
continuum images were made with the same angular resolution.  The
distribution of the continuum emission is fairly similar amongst the
four frequencies. The 686~GHz emission shows the same peaks as
previously identified by \cite{saka11} at higher angular resolution,
marked with black crosses in Fig.~\ref{fig1}. We labelled them L2 to
L7 in this paper to facilitate a comparison to the line emission as
the continuum peaks agree well in position with the corresponding line
peaks (see definitions in Fig.~\ref{fig3}). L5 and L6 appear to be
slightly closer together at 686~GHz than at the lower frequencies and
L2 is somewhat offset in Declination, i.e., $\sim$2$''$ ($\simeq$half
the synthesized beam), at 686GHz than at the lower frequencies.  This
discrepancy is larger than the uncertainties in the absolute and
relative positions of the peaks and might suggest a region of warmer
dust close to L2. Indeed, \citet{brad03} identified an overabundance
of $^{12}$CO(J=7--6) in the same region as well.  The brightnesses of
the different peaks, and their respective ratios to each other are
very similar along the five continuum peaks, with L5 being the
strongest, followed by L6, L4, L7 and L2.

\begin{figure}[!t]
   \centering
   \resizebox{\hsize}{!}{\includegraphics{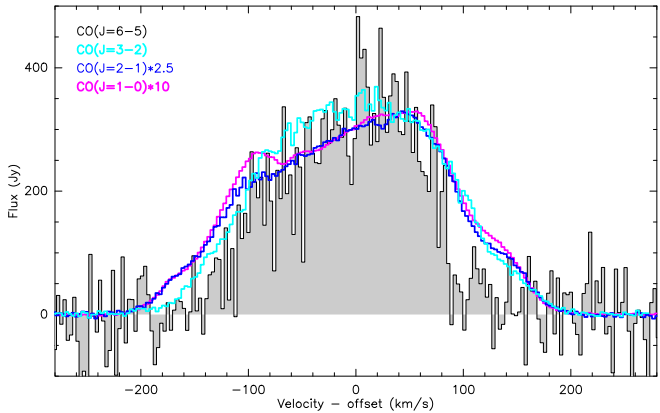}}
   \caption{The flux of the four different line transitions integrated
     over the entire emission region, respectively. It is clear from
     the broad profile that several Gaussian components are present
     for each CO transition, indicating the presence of multiple
     molecular clouds or clumps at slightly different
     velocities. However the velocity range is similar for each line,
     indicating that the gas is co-spatial in the clumps. }
   \label{fig4}
 \end{figure}

\begin{deluxetable*}{@{}l@{}@{}rrrrrrrrrrrrr}
\centering
\tablewidth{0pt}
\tablecaption{Molecular Cloud Complexes in all observed $^{12}$CO transitions.}
\tablehead{
& \multicolumn{4}{c}{\hrulefill\hskip 3mm \raisebox{-.8mm} {Bluest Gaussian Line} \hskip 3mm \hrulefill} 
& \multicolumn{4}{c}{\hrulefill\hskip 3mm \raisebox{-.8mm} {Central Gaussian Line} \hskip 3mm \hrulefill} 
& \multicolumn{4}{c}{\hrulefill\hskip 3mm \raisebox{-.8mm} {Reddest Gaussian Line} \hskip 3mm \hrulefill} \\[0.1cm]
ID & \colhead{S$_{1,\nu}^{\rm peak}$}  & \colhead{I$_{1,\nu}^{\rm peak}$} & \colhead{$v_0$}  & \colhead{$\Delta$v} 
              & \colhead{S$_{2,\nu}^{\rm peak}$}  & \colhead{I$_{2,\nu}^{\rm peak}$} & \colhead{$v_0$}  & \colhead{$\Delta$v} 
              & \colhead{S$_{3,\nu}^{\rm peak}$}  & \colhead{I$_{3,\nu}^{\rm peak}$} & \colhead{$v_0$}  & \colhead{$\Delta$v} 
 & \colhead{$\sum_{i=1}^3$(I$_{i,\nu}^{\rm peak}$)} \\
(1) & (2) & (3) & (4) & (5) & (6) & (7) & (8) & (9) & (10) & (11) & (12) & (13) &  (14) }
\startdata 
\multicolumn{14}{c}{\hrulefill\hskip 3mm \raisebox{-.8mm} \cosf \hskip 3mm \hrulefill} \\[0.1cm] 
 L0  &        --       &       --         &    -- &    -- &   $<$5.0        &  $<$426           &  -97  &   80  &  12.3$\pm$ 2.8  &    1699$\pm$ 406  &  -26  &  130  &  1699$\pm$ 406  \\
 L1  &        --       &       --         &    -- &    -- &   6.1$\pm$ 2.3  &     846$\pm$ 324  & -102  &  130  &  26.6$\pm$ 2.6  &    2860$\pm$ 398  &  -40  &  101  &  3706$\pm$ 513  \\
 L2  &  $<$5.00        &  $<$346          & -152  &   65  &  35.2$\pm$ 1.8  &    3937$\pm$ 424  &  -64  &  105  &  15.7$\pm$ 1.9  &    1402$\pm$ 240  &   38  &   84  &  5339$\pm$ 487  \\
 L3  &  $<$5.00        &  $<$378          & -129  &   71  &  27.5$\pm$ 2.7  &    4218$\pm$ 511  &  -50  &  144  &   5.0$\pm$ 2.6  &     663$\pm$ 350  &   32  &  124  &  4881$\pm$ 619  \\
 L4  &  $<$5.00        &  $<$266          & -110  &   50  &  32.0$\pm$ 2.5  &    4767$\pm$ 509  &  -52  &  140  &  22.9$\pm$ 2.4  &    2684$\pm$ 376  &   40  &  110  &  7451$\pm$ 633  \\
 L5  &  $<$5.00        &  $<$266          & -120  &   50  &  15.8$\pm$ 2.6  &    1377$\pm$ 283  &  -81  &   82  &  41.7$\pm$ 1.7  &    5327$\pm$ 496  &   22  &  120  &  6704$\pm$ 571  \\
 L6  &  $<$5.00        &  $<$160          &  -91  &   30  &  48.2$\pm$ 2.4  &    4618$\pm$ 564  &    6  &   90  &  14.6$\pm$ 2.7  &    1165$\pm$ 264  &   95  &   75  &  5783$\pm$ 623  \\
 L7  &        --       &       --         &    -- &    -- &  25.6$\pm$ 2.2  &    2450$\pm$ 342  &   12  &   90  &   $<$5.0        &     $<$426        &  112  &   80  &  2450$\pm$ 342  \\
 L8  &  $<$5.00        &  $<$378          &  -26  &   71  &   5.4$\pm$ 2.0  &     420$\pm$ 166  &   50  &   73  &   $<$5.0        &     $<$415        &  126  &   78  &   420$\pm$ 166  \\
\multicolumn{14}{c}{\hrulefill\hskip 3mm \raisebox{-.8mm} \cott \hskip 3mm \hrulefill} \\[0.1cm]                                                                                          
 L0  &        --       &       --         &    -- &    -- &   2.9$\pm$ 0.2  &     249$\pm$  37  &  -97  &   80  &   3.0$\pm$ 0.2  &     408$\pm$  40  &  -26  &  130  &   657$\pm$  54  \\
 L1  &        --       &       --         &    -- &    -- &   1.5$\pm$ 0.2  &     207$\pm$  39  & -102  &  130  &  15.4$\pm$ 0.3  &    1654$\pm$ 167  &  -40  &  101  &  1861$\pm$ 171  \\
 L2  &  2.45$\pm$ 0.2  &     169$\pm$ 31  & -152  &   65  &  13.9$\pm$ 0.2  &    1559$\pm$ 150  &  -64  &  105  &   7.1$\pm$ 0.2  &     635$\pm$  78  &   38  &   84  &  2363$\pm$ 172  \\
 L3  &  1.96$\pm$ 0.4  &     148$\pm$ 34  & -129  &   71  &  11.8$\pm$ 0.3  &    1811$\pm$ 135  &  -50  &  144  &   2.9$\pm$ 0.3  &     386$\pm$  51  &   32  &  124  &  2345$\pm$ 148  \\
 L4  &  3.81$\pm$ 0.4  &     203$\pm$ 47  & -110  &   50  &  13.0$\pm$ 0.3  &    1938$\pm$ 145  &  -52  &  140  &   4.5$\pm$ 0.3  &     525$\pm$  59  &   40  &  110  &  2666$\pm$ 163  \\
 L5  &  1.45$\pm$ 0.4  &      77$\pm$ 25  & -120  &   50  &   9.3$\pm$ 0.3  &     815$\pm$ 103  &  -81  &   82  &  16.4$\pm$ 0.2  &    2092$\pm$ 176  &   22  &  120  &  2984$\pm$ 205  \\
 L6  &  2.44$\pm$ 0.6  &      78$\pm$ 32  &  -91  &   30  &  21.3$\pm$ 0.3  &    2042$\pm$ 229  &    6  &   90  &   8.9$\pm$ 0.4  &     714$\pm$ 100  &   95  &   75  &  2834$\pm$ 252  \\
 L7  &        --       &       --         &    -- &    -- &  16.5$\pm$ 0.2  &    1583$\pm$ 177  &   12  &   90  &   5.4$\pm$ 0.2  &     457$\pm$  59  &  112  &   80  &  2040$\pm$ 187  \\
 L8  &  1.50$\pm$ 0.2  &     113$\pm$ 20  &  -26  &   71  &   8.7$\pm$ 0.2  &     678$\pm$  94  &   50  &   73  &   2.7$\pm$ 0.2  &     222$\pm$  31  &  126  &   78  &  1013$\pm$ 101  \\
\multicolumn{14}{c}{\hrulefill\hskip 3mm \raisebox{-.8mm} \coto \hskip 3mm \hrulefill} \\[0.1cm]                                                                                          
 L0  &        --       &       --         &    -- &    -- &   4.2$\pm$ 0.1  &     360$\pm$  46  &  -97  &   80  &   2.1$\pm$ 0.1  &     291$\pm$  24  &  -26  &  130  &   651$\pm$  52  \\
 L1  &        --       &       --         &    -- &    -- &   2.8$\pm$ 0.1  &     383$\pm$  34  & -102  &  130  &  12.6$\pm$ 0.1  &    1357$\pm$ 135  &  -40  &  101  &  1740$\pm$ 139  \\
 L2  &  4.08$\pm$ 0.1  &     282$\pm$ 44  & -152  &   65  &  12.2$\pm$ 0.1  &    1362$\pm$ 130  &  -64  &  105  &   5.6$\pm$ 0.1  &     496$\pm$  60  &   38  &   84  &  2140$\pm$ 150  \\
 L3  &  4.10$\pm$ 0.2  &     310$\pm$ 46  & -129  &   71  &   8.3$\pm$ 0.2  &    1279$\pm$  92  &  -50  &  144  &   3.6$\pm$ 0.2  &     479$\pm$  44  &   32  &  124  &  2068$\pm$ 112  \\
 L4  &  4.27$\pm$ 0.2  &     227$\pm$ 47  & -110  &   50  &   9.1$\pm$ 0.2  &    1353$\pm$  99  &  -52  &  140  &   6.2$\pm$ 0.2  &     725$\pm$  68  &   40  &  110  &  2305$\pm$ 129  \\
 L5  &  3.85$\pm$ 0.2  &     205$\pm$ 43  & -120  &   50  &   5.2$\pm$ 0.2  &     454$\pm$  58  &  -81  &   82  &  13.2$\pm$ 0.1  &    1688$\pm$ 142  &   22  &  120  &  2347$\pm$ 159  \\
 L6  &  1.19$\pm$ 0.1  &      38$\pm$ 13  &  -91  &   30  &  17.7$\pm$ 0.1  &    1696$\pm$ 189  &    6  &   90  &   6.9$\pm$ 0.1  &     550$\pm$  74  &   95  &   75  &  2284$\pm$ 203  \\
 L7  &        --       &       --         &    -- &    -- &  14.2$\pm$ 0.2  &    1360$\pm$ 152  &   12  &   90  &   4.0$\pm$ 0.2  &     344$\pm$  45  &  112  &   80  &  1704$\pm$ 159  \\
 L8  &  1.25$\pm$ 0.1  &      95$\pm$ 15  &  -26  &   71  &   9.3$\pm$ 0.1  &     720$\pm$  99  &   50  &   73  &   2.6$\pm$ 0.1  &     219$\pm$  29  &  126  &   78  &  1034$\pm$ 104  \\
\multicolumn{14}{c}{\hrulefill\hskip 3mm \raisebox{-.8mm} \cooz \hskip 3mm \hrulefill} \\[0.1cm]                                                                                          
 L0  &       --     -- &       --         &    -- &    -- &  0.30$\pm$0.01  &      25$\pm$   3  &  -97  &   80  &  0.03$\pm$0.01  &       5$\pm$   2  &  -26  &  130  &    30$\pm$   4  \\
 L1  &       --     -- &       --         &    -- &    -- &  0.16$\pm$0.02  &      22$\pm$   3  & -102  &  130  &  1.84$\pm$0.02  &     197$\pm$  20  &  -40  &  101  &   219$\pm$  20  \\
 L2  &  0.34$\pm$0.04  &      23$\pm$  4  & -152  &   65  &  1.66$\pm$0.03  &     185$\pm$  18  &  -64  &  105  &  0.44$\pm$0.03  &      40$\pm$   6  &   38  &   84  &   248$\pm$  19  \\
 L3  &  0.42$\pm$0.04  &      32$\pm$  5  & -129  &   71  &  1.11$\pm$0.03  &     171$\pm$  13  &  -50  &  144  &  0.17$\pm$0.03  &      22$\pm$   4  &   32  &  124  &   225$\pm$  14  \\
 L4  &  0.63$\pm$0.06  &      34$\pm$  8  & -110  &   50  &  1.45$\pm$0.04  &     216$\pm$  17  &  -52  &  140  &  0.58$\pm$0.04  &      68$\pm$   8  &   40  &  110  &   318$\pm$  20  \\
 L5  &  0.43$\pm$0.05  &      23$\pm$  5  & -120  &   50  &  0.77$\pm$0.04  &      67$\pm$   9  &  -81  &   82  &  1.67$\pm$0.03  &     213$\pm$  18  &   22  &  120  &   303$\pm$  21  \\
 L6  &  0.06$\pm$0.04  &       2$\pm$  1  &  -91  &   30  &  2.57$\pm$0.04  &     246$\pm$  28  &    6  &   90  &  0.95$\pm$0.04  &      76$\pm$  11  &   95  &   75  &   322$\pm$  30  \\
 L7  &       --     -- &       --         &    -- &    -- &  1.96$\pm$0.03  &     188$\pm$  21  &   12  &   90  &  0.51$\pm$0.03  &      43$\pm$   6  &  112  &   80  &   231$\pm$  22  \\
 L8  &  0.08$\pm$0.01  &       6$\pm$  1  &  -26  &   71  &  1.20$\pm$0.01  &      93$\pm$  13  &   50  &   73  &  0.31$\pm$0.01  &      26$\pm$   4  &  126  &   78  &   125$\pm$  14  \\

\enddata

\tablenotetext{(1)}{Names of \cosf\, line emission peaks as given in
  this paper (starting with an L). Corresponding \cott\, and \coto\,
  names from \cite{saka11}: L0=SB2; L2$\simeq$Sa1; L4$\simeq$Sa2;
  L5$\simeq$Sa3; L6$\simeq$Sa4; L7$\simeq$Sa5, L8=SB1. The absolute
  positions of each of these line peaks are (for $\alpha$=00$^{\rm
    h}$47$^{\rm m}$, $\delta$=$-25^\circ17'$): L0=34.56s,09.2$"$;
  L1=34.07s,12.0$"$; L2=33.69s,14.2$"$; L3=33.54s,14.5$"$; L4=33.31s,15.8$"$;
  L5=33.15s,17.5$"$; L6=32.97s,19.4$"$; L7=32.80s,21.7$"$; L8=32.53s,24.0$"$. }

\tablenotetext{(2,3,6,7,10,11)}{Peak flux densities S$_\nu$ (in
  Jy\,beam$^{-1}$) and spectrally integrated intensities I$_{i,\nu}$
  (=$\int\left(S_{i,\nu}^{\rm int}~dv\right)$) in
  (Jy\,beam$^{-1}$\,\kms) derived at the peak emission for each fitted
  Gaussian line.}

\tablenotetext{(4,8,12)}{Zero velocity is with respect to the
  redshifted frequency of the observed CO line transition. The central
  velocities of each Gaussian line component have been first derived
  by fitting the \coto\, and \cott\, lines. The so found values have
  been averaged and subsequently fixed for each fit. We assume a
  5-10~\kms\ error on the central velocities. The unit is \kms.}

\tablenotetext{(5,9,13)}{Full width at half maximum (FWHM) of each
  line. Identical to the velocity centre, the FWHM have been first
  derived by fitting the \coto\, and \cott\, lines, taking the mean of
  both and subsequently fixing it in the fits for all line
  transition. We assume an error on the FWHM of 5-10\kms. The unit is
  in \kms.}

\label{tab3}
\end{deluxetable*}

Table~\ref{tab2} lists the total continuum fluxes for all four
frequencies and Table~\ref{tab2.5} the individual fluxes per beam for
each continuum peak. Assuming no significant resolution effects, the
different fluxes indicate a total spectral index of $+$3 (for
$S_\nu\propto\nu^\alpha$), see also Fig.~\ref{fig2}. This is
consistent with the spectral index found between the 1.3~mm and
0.87~mm continuum emission by \citet{saka11} and between the 2.6~mm
and 1.3~mm continuum by \citet{saka06}. This strongly indicates that
the emission at $>$115~GHz is largely dominated by thermal dust
emission.  Our new 690~GHz data, along with the ALMA 115~GHz data, now
reveal differences in continuum intensity ratios among the continuum
peaks (see Fig.~\ref{fig2} and Fig.~\ref{fig2.5}); Fig.~\ref{fig2}
shows the spectral emissitivity index maps between the different
frequencies and Fig.~\ref{fig2.5} plots the continuum SED for each of
the continuum peaks normalized to the peak flux at 220~GHz for each
component. While it is true that along L4-L7 the ratios do not change
too much, yielding still an index around 3, at least for the higher
frequencies, it increases to $\sim$4-5 at L2 (see
Fig.~\ref{fig2}). This appears to be consistent with the overabundance
of $^{12}$CO(J=7--6) found by \citet{brad03} suggesting a warmer dust
component in this region.  The latter would add to a higher flux of
the 690~GHz continuum emission assuming a two temperature phase of the
dust here.  Also, we see a slight drop of the spectral index with
respect to the 2.6~mm continuum which most likely indicates the
increasing role of non-thermal synchrotron and thermal free-free
continuum emission towards longer wavelengths (see Fig.~\ref{fig2.5});
\cite{hees11} and \cite{ulv97} find a probably equal contribution
between non-thermal synchrotron radio emission presumably from
supernova remnants and thermal free-free radio emission from HII
regions, i.e., star formation. \cite{hees11} derive a spectral index
of $\sim$ $-0.5$ for the radio continuum emission at cm wavelengths in
the center which steepens to $-2$ toward the outer parts and
filaments.  Judging from Figure~9 in \cite{hees11}, L5 and L6 show a
spectral index of around $-$0.5, while L2, L4 and L7 increase to
around $-$1, explaining the differences in the deviations of the
115~GHz continuum from the thermal (sub)mm (modified black body) slope
for the different components. Depending on the component and assumed
spectral index, the synchrotron/free-free emission contributes between
15\% and 80\% to the 115~GHz continuum emission, with L2 having the
least (15\%) and L5 and L7 ($\sim$80\%) having the most contribution
from these radio (cm/mm) processes to their 2.6~mm continuum
emission. However, if the synchrotron/free-free spectral indices are
somewhat flatter or steeper, the fraction of this emission to the
total continuum emission will change accordingly (being higher or
lower, respectively). Given that generally (but not exclusively) the
radio spectra of non-thermal synchrotron continuum emission is steeper
(i.e., $\alpha\simeq-0.5$ to $-1$) than that of the thermal free-free
continuum emission, it might be likely that the continuum emission at
2.6~mm has a larger contribution of thermal free-free emission
\cite[see also][]{peel2011}. However, this might change from clump to
clump within the disk and we have no actual way to distinguish between
non-thermal synchrotron and thermal free-free emission from our
continuum observations other than the spectral radio index (which
might be misleading). Therefore, we will not further stress a
differentiation between the two processes.

\subsubsection{Temperature, opacity and mass of the dust}
\label{dustmass}

While fitting a simple power law to the (sub-)millimeter dust
  emission is a reasonable first-order approach, we can actually do
  better and determine the temperature, opacity and dust mass from our
  observations, using standard equations \cite[for a comparison on
  similar observations and sources see also the recent work on 690~GHz
  ALMA data for Arp220 by][]{wilson2014,scoville2015}. By using flux
  ratios, we can eliminate some assumptions such as on the dust
  emissivity, in particular to derive the dust opacities and dust
  temperatures (but see also further below for the dust mass for which
  this approach does not work). This gives the following equation
  \cite[see also][]{wilson2014}:

  \begin{equation}
   \frac{\rm S_{\nu_1}}{\rm S_{\nu_2}} = \left(\frac{\nu_1}{\nu_2}\right)^3
    \left(\frac{e^{h\nu_2/kT_d}-1}{e^{h\nu_1/kT_d}-1}\right)
    \left(\frac{1-e^{-\tau_1}}{1-e^{-\tau_2}}\right)
  \end{equation}

  Using the flux densities of the emission at
  220~GHz\footnote{Contributions from the free-free/synchrotron
    emission can be neglected here as they are less than 5\% being
    hence smaller than the uncertainties and leaving unchanged the
    final best fit solutions.}, 350~GHz and 686~GHz, this allows us to
  derive the opacities and dust temperatures in the different peaks,
  as shown in Fig.\ref{fig2.5} along with the $\chi^2$-fits. We
  assumed a spectral index of the dust emissivity\footnote{We fitted
    equally for a spectral index of 1.5 and 2.1 as used in previous
    publications for NGC~253 \cite[e.g.,][]{peel2011}, but the best
    fit solutions were found for a value of 1.8} of 1.8 \cite[see][and
  references therein]{wilson2014} and varied the dust temperatures in
  the range of 2.5~K up to 150~K and opacities at 220~GHz between 0.02
  and 1 (corresponding to opacities at 686~GHz of
  $\tau_{686}=(\nu_{686}/\nu_{220})^{1.8}\cdot\tau_{220}$, i.e.,
  $\sim$0.15 to 8). The best-fit solutions suggest dust temperatures
  of $\sim$10-30~K at low opacities of $\tau_{686}$=0.19 for all five
  peaks, with L2 having the highest T$_{\rm d}$ of 25~K. However, L4
  and L5 have almost equally low-$\chi^2$ solutions at higher T$_{\rm
    d}>60~K$ and much higher opacities of $\tau_{686}$=4 while the
  opacities of L2 stay consistently below 1. The two-temperature
  best-fit ranges are in good agreement with what is found for the
  molecular gas that can only be well fit with a two-phase temperature
  gas model (see next Sections). As dust and molecular gas are
  intimately bound, one would expect a two-temperature phase as well
  in the dust. The higher T$_{\rm d}$ of the inner disk are similar to
  the values found for the eastern nucleus in Arp~220 which was fitted
  to $\sim$80~K but at higher opacities that are closer to the values
  found in the western nucleus of Arp~220 \cite[][]{wilson2014}.

  To calculate the dust mass M$_{\rm dust}$ at a given frequency
  for each peak, we use the following standard equation:\\
    \begin{equation}
      {\rm M}_{\rm dust} = \frac{{\rm S}_\nu{\rm D}^2}{\kappa_\nu{\rm B}(\nu,{\rm T}_{\rm d})}
    \end{equation}
    with S$_\nu$ being the flux observed at frequency $\nu$, D the
    distance (=3.5~Mpc), $\kappa_\nu$ the mass absorption coefficient
    $\kappa_\nu=\kappa_0(\nu/\nu_0)^\beta$ with $\beta$ being the
    spectral index of the dust emissivity, and B($\nu$,T$_{\rm d}$)
    the Planck function. Assuming $\beta$=1.8 and $\kappa_{158\mu\rm
      m}\simeq$0.6-3~m$^2$/g at a frequency of $\nu_0$=1.9~THz
    ($\equiv$158$\mu m$) depending on the dust properties
    \cite[see][]{hirash2014,dayal2010,zubko1996,zubko2004,draine1984},
    we estimate dust masses of $\sim$10$^{4}$-10$^{5}$-M$_\odot$ per
    dust peak, i.e. with an uncertainty of a factor of 5-10. The
    biggest uncertainty comes from the range of $\kappa_\nu$ that
    depends on which type or combination of dust grains dominates
    \cite[i.e., graphite, silicate, amorphous carbon, etc., see for a
    summary][]{hirash2014}. Comparing the dust masses to the molecular
    gas masses derived from optically thin molecular tracers (such as
    $^{13}$CO) of a few $\sim$10$^{7}$~M$_\odot$ \cite[see for
    instance][]{leroy15}, we derive a gas-to-dust mass ratio of
    $\gtrsim$100-1000 for the different peaks and the range in
    $\kappa$.  Doing the same exercise on the global flux of the
    entire disk, we find a dust mass of about a few 10$^{6}$M$_\odot$
    for a dust temperature of T$_{\rm d}$=25~K. The gas mass of the
    disk in NGC253 is about a few 10$^8$-10$^9$~M$_\odot$ in the area
    of the detected continuum emission, resulting in a global
    gas-to-dust mass ratio of a few hundred. The gas-to-dust mass
    ratios are consistent with the values ($\sim$100-1000) usually
    found in active galaxies with metallicities around solar \cite[see
    for instance][]{remy2014}.

    The global (and individual) dust temperatures found here are in
    good agreement with previous works on NGC~253 \cite[see for
    instance][]{peel2011}.  The global dust mass of NGC~253 is similar
    to that found in nearby active galaxies \cite[see for
    instance][]{remy2014} but one to two orders of magnitude smaller
    than that found for nearby (ultra-)luminous-infrared galaxies
    (=(U)LIRGs) such as Arp~220, NGC~6240 or NGC~1614 \cite[see for
    instance][]{scoville2015,xu15}.  Arp~220 and NGC~6240 show dust
    masses around a few 10$^9$M$_\odot$ and NGC~1614 has a dust mass
    around 10$^{7}$-10$^8$M$_\odot$. This is not surprising as
    especially the nuclei in Arp~220 are known to be highly obscured
    exhibiting a starburst much stronger than that in the disk of
    NGC~253 while NGC~1614 is likely more comparable to NGC~253 than
    Arp~220.

\subsection{Line Emission}

\begin{figure*}[!t]
   \centering
   \resizebox{\hsize}{!}{\includegraphics{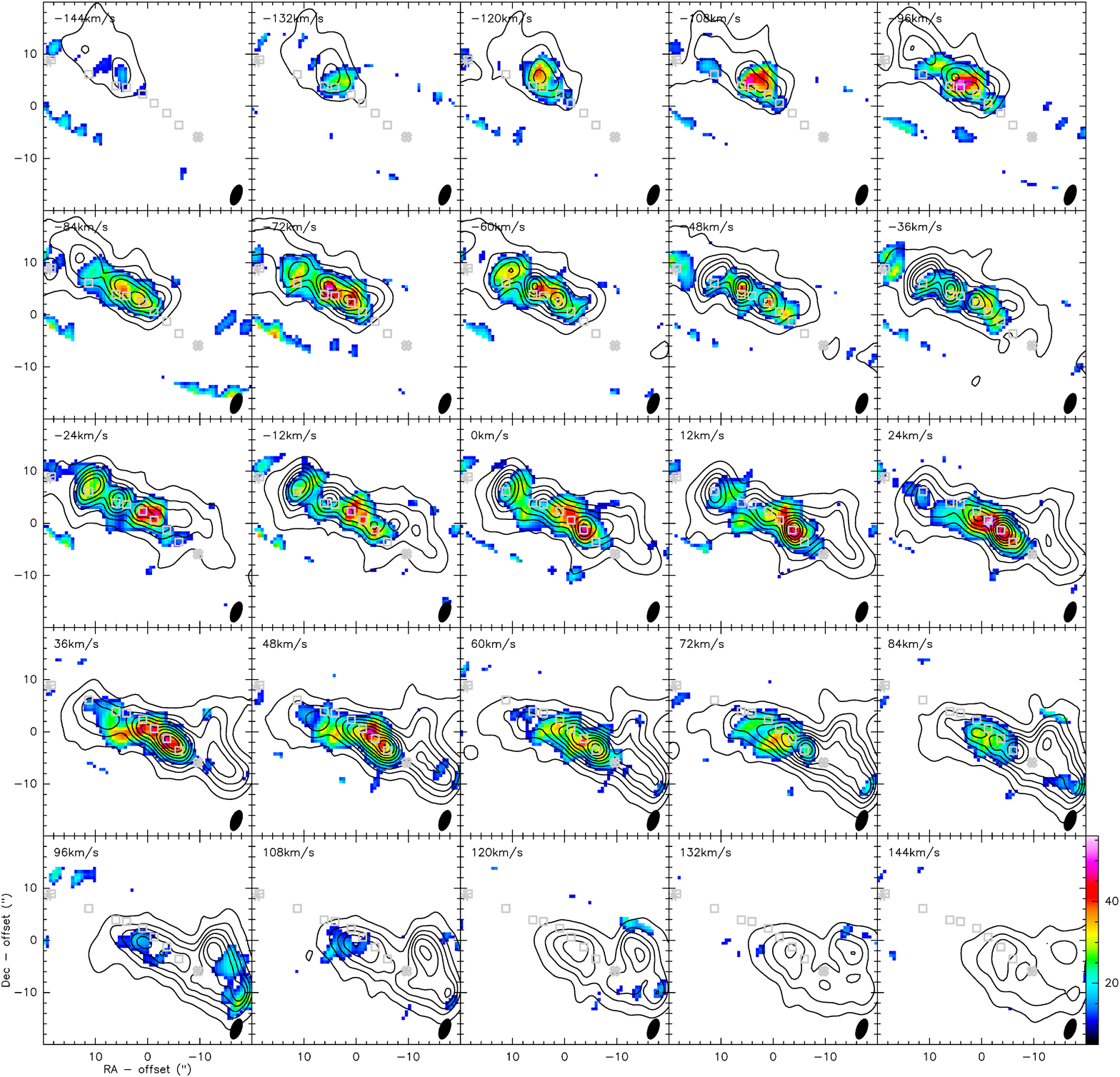}}
   \caption{Channel maps of the \cooz\, (black contours) overlaid onto
     the \cosf\, emission (color scale). The spectral resolution is
     $\sim$12~\kms. The contours for the \cooz\, emission start at
     100$\sigma$=0.1~Jy~beam$^{-1}$~channel$^{-1}$ and go in steps of
     300$\sigma$. The grey symbols represent the seven different line
     peaks in the disk and the stars the two shells as shown in
     Fig.~\ref{fig3} and specified in Table~\ref{tab3}.}
   \label{fig5.0}
\end{figure*}

\subsubsection{General Properties}

We clearly detect emission from \cosf\, (Fig.~\ref{fig3}). The
distribution of the spectrally integrated \cosf\, line emission
closely follows that of \cott, \coto\, and \cooz\, (see
Fig.~\ref{fig3}).  We identify at least seven different peaks in the
integrated \cosf\, emission, labeled L1 to L7 in Fig.~\ref{fig3}, plus
the location of the two shells labeled L0 (for SB2) and L8 (for
SB1). L4 to L6 appear to be the most dominant ones in all four
$^{12}$CO transitions. L5 is spatially coincident with the dynamical
center of NGC~253 \citep[see][]{ros12,muel10}. Most of these line
peaks in the disk, with the exceptions of L1 and L3 which have no
correspondence in the continuum emission, spatially coincide with the
continuum peaks at 2.6~mm, 1.3~mm, 0.87~mm and 0.43~mm identified in
\citet{saka11} and this work (see also Section~\ref{seccont}). Line
peaks L4 and L5 appear merged for the \cott\, and \coto\ emission at
the angular resolution used in this paper while they are clearly
separated for the \cosf\, emission. However, at the higher angular
resolution used in \citet{saka11}, the peak positions match well those
seen for \cosf, with the exception of L3, which does not have an
obvious counterpart in the lower transitions. This merging of the
peaks is probably due to extended emission in the lower $^{12}$CO
transitions as \cosf\, most likely traces denser (and/or hotter)
gas. L5 appears to be the strongest peak for all lines except \cooz,
where L6 seems to dominate.  Interestingly, L4 appears to be somewhat
brighter in \cosf, \coto\, and \cooz\, than in \cott, an effect that
is already seen for the $^{13}$CO(J=2--1), C$^{18}$O(J=2--1) and
HCN(J=4--3) emission by \citet{saka11}.

We only detect marginal \cosf\, emission around the western and
eastern shells (SB1=L08 and SB2=L01) described in \citet{saka06} and
marked with stars and squares in Fig.~\ref{fig3}. While the eastern
shell is only partially covered with our five-point mosaic of the
\cosf\, emission, the western shell is (almost) completely
covered. However, the western shell is already significantly weaker in
the lower $^{12}$CO transitions, so that we might still simply lack
adequate sensitivity in our observations for a clear detection.

\begin{figure*}[!t]
   \centering
   \resizebox{\hsize}{!}{\includegraphics{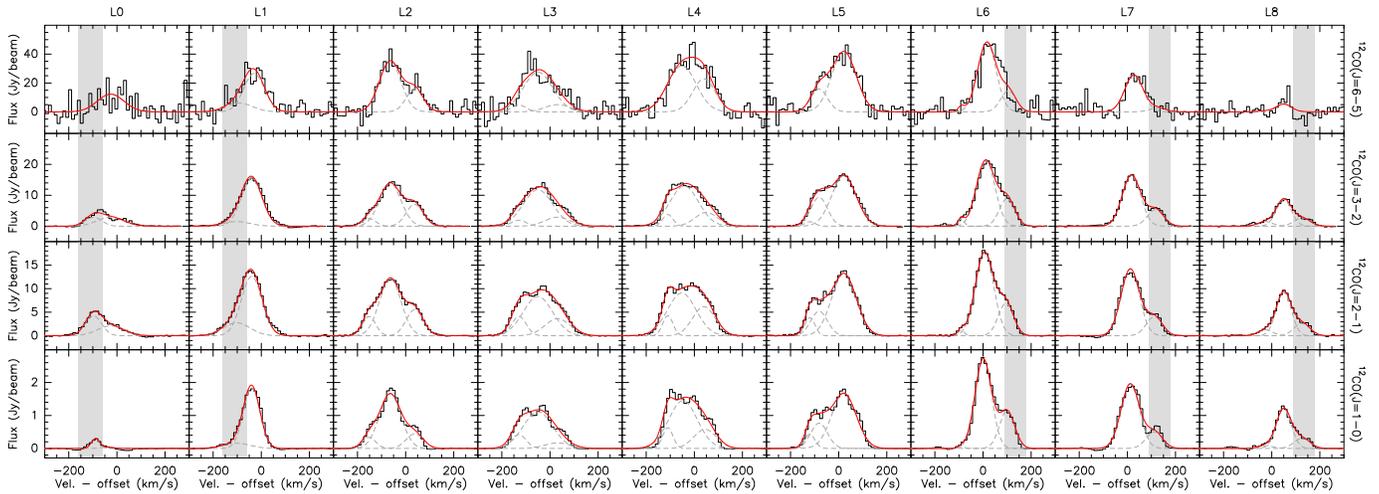}}
   \caption{Line spectra for the different line peaks for each
     $^{12}$CO transition.  Each row corresponds to one transition
     (see labelling on the right) and each column to the different
     line peaks L0 to L8 (see labelling on the top). A Gaussian fit
     with up to three components has been applied to all spectra,
     shown as solid red ({\it sum}) and dashed grey curves ({\it
       individual components}). We fixed the velocities and FWHM
     during the fit to ensure that we fit the same components in all
     four transitions. The results of these fits are given in
     Table~\ref{tab2} and \ref{tab3}. The grey shaded areas
     approximately mark the velocity ranges for the eastern (around
     L0) and western shell (around L8 and close to L6/L7) in the wings
     of the line profiles. }
   \label{fig8.0}
\end{figure*}

The line profiles of the \cosf, \cott, \coto\, and \cooz\, emission,
shown in Fig.~\ref{fig4}, were spatially integrated over the entire
emission area for each transition respectively.  Multiple-component
(up to three) Gaussian fits were carried out on the line profiles and
the results are given in Table~\ref{tab2}. As can be seen, while
overall the line profiles agree well with each other, part of the
redshifted and blueshifted wing of the \cott, \coto\, and \cooz\, line
emission is apparently missing in the \cosf\, line emission. These
``wing''-components are most likely associated with the western
(redshifted velocities) and eastern (blueshifted velocities) shells
\citep[see][]{saka06}. While the eastern part could easily still be
hidden in the noise, one gets the impression that the redshifted
\cosf\, emission is indeed not present and that this is more likely
due to excitation effects than to insufficient sensitivity as also
shown by the individual spectra at the peaks (see Fig.~\ref{fig8.0}).

Looking at the velocity channel maps in Fig.~\ref{fig5.0}, in which
the \cosf\, emission is exemplary plotted in contours over the \cooz\,
emission in color scale, we find again that all four transitions
strongly resemble each other for most velocities; especially the three
lowest transitions are almost indistinguishable from each other (see
also Fig.~\ref{fig4} \& \ref{fig8.0}). Although it is difficult to
clearly identify the line peaks (Fig.~\ref{fig3}) of the \cosf\,
emission in the velocity resolved channels due to the lower
signal-to-noise ratios and because each of these peaks represent an
ensemble of giant molecular cloud complexes (GMCs), one can still see
some trends. L6 clearly is the dominant and brightest peak for the
\cott, \coto\, and \cooz\, emission between velocities 24~\kms\, and
60~\kms\, with L4 and L5 being significantly weaker as opposed to the
spectrally integrated maps in Fig.~\ref{fig3}. At these velocities,
both L5 and L6 appear almost equally bright in the \cosf\,
emission. L3 and L4 also look equally strong in \cosf\, for velocities
around $-$80~\kms and $-$48~\kms, while L3 seems to be brighter for
the lower transitions.  The lack of significant \cosf\, emission at
the higher velocities ($\gtrsim$100~\kms and $\lesssim-$100~\kms) with
respect to the lower three transitions is again striking.

\begin{deluxetable*}{lrrrrrrrrr}[!t]
\centering
\tablewidth{0pt}
\tablecaption{$^{12}$CO brightness temperature ratios from individual Gaussian Components}
\tablehead{
\colhead{Ratio} & \colhead{L0} & \colhead{L1} & \colhead{L2} & \colhead{L3} 
& \colhead{L4} & \colhead{L5} & \colhead{L6} & \colhead{L7} & \colhead{L8} \\
     (1) & (2)            &  (3)	   & (4)    & (5)  & (6)  & (7)     
         & (8)            &  (9)           & (10)   }
\startdata 							  				

\multicolumn{10}{c}{\hrulefill\hskip 3mm \raisebox{-.8mm}{Bluest Gaussian Line} \hskip 3mm \hrulefill} \\[0.2cm] 
R$_{65}^{32}$                  &       --       &       --       & $>$2.00        & $>$1.60        & $>$3.10        & $>$1.20        & $>$1.90        &       --      &  $>$1.20       \\[0.1cm]
R$_{65}^{21}$                  &       --       &       --       & $>$7.30        & $>$7.40        & $>$7.70        & $>$6.90        & $>$2.10        &       --      &  $>$2.30       \\[0.1cm]
R$_{65}^{10}$                  &       --       &       --       & $>$2.40        & $>$3.00        & $>$4.60        & $>$3.10        & $>$0.4         &       --      &  $>$0.60       \\[0.1cm]
R$_{32}^{21}$                  &       --       &       --       & 3.80$\pm$0.90  & 4.70$\pm$1.30  & 2.50$\pm$0.80  & 6.00$\pm$2.30  & 1.10$\pm$0.60  &       --      &  1.90$\pm$0.40 \\[0.1cm]
R$_{32}^{10}$                  &       --       &       --       & 1.20$\pm$0.30  & 1.90$\pm$0.50  & 1.50$\pm$0.50  & 2.70$\pm$1.10  & 0.20$\pm$0.10  &       --      &  0.50$\pm$0.10 \\[0.1cm]
R$_{21}^{10}$                  &       --       &       --       & 0.33$\pm$0.08  & 0.41$\pm$0.09  & 0.60$\pm$0.19  & 0.45$\pm$0.14  & 0.21$\pm$0.13  &       --      &  0.25$\pm$0.06 \\[0.3cm]
${\rm R}_{65}^{32}/R_{\rm L4}$   &      --        &      --        &  --            &  --            &  --            &  --            &  --            &      --       &   --           \\[0.1cm]
${\rm R}_{65}^{21}/R_{\rm L4}$   &      --        &      --        &  --            &  --            &  --            &  --            &  --            &      --       &   --           \\[0.1cm]
${\rm R}_{65}^{10}/R_{\rm L4}$   &      --        &      --        &  --            &  --            &  --            &  --            &  --            &      --       &   --           \\[0.1cm]
${\rm R}_{32}^{21}/R_{\rm L4}$   &      --        &      --        &  1.5$\pm$0.6   &  1.9$\pm$0.8   &  1.0           &  2.4$\pm$1.2   &  0.4$\pm$0.2   &      --       &   0.8$\pm$0.3  \\[0.1cm]
${\rm R}_{32}^{10}/R_{\rm L4}$   &      --        &      --        &  0.8$\pm$0.3   &  1.3$\pm$0.6   &  1.0           &  1.8$\pm$0.9   &  0.2$\pm$0.1   &      --       &   0.3$\pm$0.1  \\[0.1cm]
${\rm R}_{21}^{10}/R_{\rm L4}$   &      --        &      --        &  0.5$\pm$0.2   &  0.7$\pm$0.3   &  1.0           &  0.7$\pm$0.3   &  0.4$\pm$0.3   &      --       &   0.4$\pm$0.2  \\[0.2cm]
\multicolumn{10}{c}{\hrulefill\hskip 3mm \raisebox{-.8mm}{Central Gaussian Line} \hskip 3mm \hrulefill} \\[0.2cm] 
R$_{65}^{32}$                  & $>$2.30        & 1.00$\pm$0.40  & 1.60$\pm$0.20  & 1.70$\pm$0.20  & 1.60$\pm$0.20  & 2.40$\pm$0.60  & 1.80$\pm$0.30  & 2.60$\pm$0.50 &  6.50$\pm$2.70 \\[0.1cm]
R$_{65}^{21}$                  & $>$7.60        & 4.10$\pm$1.60  & 3.10$\pm$0.40  & 2.70$\pm$0.40  & 2.60$\pm$0.30  & 3.00$\pm$0.70  & 3.30$\pm$0.50  & 5.00$\pm$0.90 & 15.40$\pm$6.50 \\[0.1cm]
R$_{65}^{10}$                  & $>$2.10        & 0.90$\pm$0.40  & 1.70$\pm$0.20  & 1.50$\pm$0.20  & 1.60$\pm$0.20  & 1.80$\pm$0.40  & 1.90$\pm$0.30  & 2.80$\pm$0.50 &  8.00$\pm$3.30 \\[0.1cm]
R$_{32}^{21}$                  & 3.30$\pm$0.60  & 4.20$\pm$0.90  & 2.00$\pm$0.30  & 1.60$\pm$0.20  & 1.60$\pm$0.20  & 1.30$\pm$0.20  & 1.90$\pm$0.30  & 1.90$\pm$0.30 &  2.40$\pm$0.50 \\[0.1cm]
R$_{32}^{10}$                  & 0.70$\pm$0.10  & 1.00$\pm$0.20  & 1.10$\pm$0.10  & 0.80$\pm$0.10  & 1.00$\pm$0.10  & 0.70$\pm$0.10  & 1.10$\pm$0.20  & 1.10$\pm$0.20 &  1.20$\pm$0.20 \\[0.1cm]
R$_{21}^{10}$                  & 0.20$\pm$0.04  & 0.23$\pm$0.04  & 0.54$\pm$0.07  & 0.53$\pm$0.06  & 0.64$\pm$0.07  & 0.59$\pm$0.11  & 0.58$\pm$0.09  & 0.55$\pm$0.09 &  0.52$\pm$0.10 \\[0.3cm]
${\rm R}_{65}^{32}/R_{\rm L4}$   & $>$1.4         &  0.6$\pm$0.3   &  1.0$\pm$0.2   &  1.1$\pm$0.2   &  1.0           &  1.5$\pm$0.4   &  1.1$\pm$0.2   &  1.6$\pm$0.4  &   4.0$\pm$1.8  \\[0.1cm]
${\rm R}_{65}^{21}/R_{\rm L4}$   & $>$3.0         &  1.6$\pm$0.7   &  1.2$\pm$0.2   &  1.1$\pm$0.2   &  1.0           &  1.2$\pm$0.3   &  1.3$\pm$0.3   &  2.0$\pm$0.4  &   6.0$\pm$2.6  \\[0.1cm]
${\rm R}_{65}^{10}/R_{\rm L4}$   & $>$1.3         &      --        &  1.0$\pm$0.2   &  0.9$\pm$0.2   &  1.0           &  1.1$\pm$0.3   &  1.2$\pm$0.3   &  1.7$\pm$0.4  &   4.9$\pm$2.2  \\[0.1cm]
${\rm R}_{32}^{21}/R_{\rm L4}$   &  2.1$\pm$0.5   &  2.7$\pm$0.6   &  1.3$\pm$0.2   &  1.0$\pm$0.1   &  1.0           &  0.8$\pm$0.2   &  1.2$\pm$0.2   &  1.2$\pm$0.2  &   1.5$\pm$0.3  \\[0.1cm]
${\rm R}_{32}^{10}/R_{\rm L4}$   &  0.6$\pm$0.1   &  1.0$\pm$0.3   &  1.1$\pm$0.2   &  0.8$\pm$0.1   &  1.0           &  0.7$\pm$0.1   &  1.1$\pm$0.2   &  1.1$\pm$0.2  &   1.2$\pm$0.3  \\[0.1cm]
${\rm R}_{21}^{10}/R_{\rm L4}$   &  0.3$\pm$0.1   &  0.4$\pm$0.1   &  0.9$\pm$0.2   &  0.8$\pm$0.1   &  1.0           &  0.9$\pm$0.2   &  0.9$\pm$0.2   &  0.9$\pm$0.2  &   0.8$\pm$0.2  \\[0.2cm]
\multicolumn{10}{c}{\hrulefill\hskip 3mm \raisebox{-.8mm}{Reddest Gaussian Line} \hskip 3mm \hrulefill} \\[0.2cm] 
R$_{65}^{32}$                 & 1.00$\pm$0.20  & 2.30$\pm$0.40  & 1.80$\pm$0.40  & 2.30$\pm$1.30  & 0.80$\pm$0.10  & 1.60$\pm$0.20  & 2.50$\pm$0.70  & $>$4.30       &  $>$2.10       \\[0.1cm]
R$_{65}^{21}$                 & 1.50$\pm$0.40  & 4.30$\pm$0.70  & 3.20$\pm$0.70  & 6.50$\pm$3.50  & 2.40$\pm$0.40  & 2.90$\pm$0.40  & 4.20$\pm$1.10  & $>$7.30       &  $>$4.70       \\[0.1cm]
R$_{65}^{10}$                 & 0.11$\pm$0.05  & 2.50$\pm$0.40  & 1.00$\pm$0.20  & 1.20$\pm$0.70  & 0.90$\pm$0.20  & 1.40$\pm$0.20  & 2.30$\pm$0.60  & $>$3.60       &  $>$2.30       \\[0.1cm]
R$_{32}^{21}$                 & 1.60$\pm$0.20  & 1.80$\pm$0.30  & 1.80$\pm$0.30  & 2.80$\pm$0.40  & 3.10$\pm$0.50  & 1.80$\pm$0.20  & 1.70$\pm$0.30  & 1.70$\pm$0.30 &  2.20$\pm$0.40 \\[0.1cm]
R$_{32}^{10}$                 & 0.11$\pm$0.04  & 1.10$\pm$0.20  & 0.60$\pm$0.10  & 0.50$\pm$0.10  & 1.20$\pm$0.20  & 0.90$\pm$0.10  & 1.00$\pm$0.20  & 0.80$\pm$0.20 &  1.10$\pm$0.20 \\[0.1cm]
R$_{21}^{10}$                 & 0.07$\pm$0.03  & 0.58$\pm$0.08  & 0.32$\pm$0.06  & 0.18$\pm$0.04  & 0.38$\pm$0.06  & 0.50$\pm$0.06  & 0.55$\pm$0.11  & 0.50$\pm$0.10 &  0.47$\pm$0.10 \\[0.3cm]
${\rm R}_{65}^{32}/R_{\rm L4}$   &  1.2$\pm$0.40  &  3.0$\pm$0.7   &  2.3$\pm$0.6   &  3.0$\pm$1.7   &  1.0           &  2.0$\pm$0.4   &  3.1$\pm$1.0   &  $>$5.5       &   $>$2.7       \\[0.1cm]
${\rm R}_{65}^{21}/R_{\rm L4}$   &  0.6$\pm$0.20  &  1.8$\pm$0.4   &  1.3$\pm$0.3   &  2.7$\pm$1.5   &  1.0           &  1.2$\pm$0.3   &  1.7$\pm$0.5   &  $>$3.0       &   $>$2.0       \\[0.1cm]
${\rm R}_{65}^{10}/R_{\rm L4}$   &  0.1$\pm$0.05  &  2.7$\pm$0.7   &  1.1$\pm$0.3   &  1.3$\pm$0.8   &  1.0           &  1.6$\pm$0.4   &  2.6$\pm$0.8   &  $>$4.0       &   $>$2.5       \\[0.1cm]
${\rm R}_{32}^{21}/R_{\rm L4}$   &  0.5$\pm$0.10  &  0.6$\pm$0.1   &  0.6$\pm$0.1   &  0.9$\pm$0.2   &  1.0           &  0.6$\pm$0.1   &  0.6$\pm$0.1   &  0.5$\pm$0.1  &   0.7$\pm$0.2  \\[0.1cm]
${\rm R}_{32}^{10}/R_{\rm L4}$   &  0.1$\pm$0.04  &  0.9$\pm$0.2   &  0.5$\pm$0.1   &  0.4$\pm$0.1   &  1.0           &  0.8$\pm$0.2   &  0.8$\pm$0.2   &  0.7$\pm$0.2  &   0.9$\pm$0.2  \\[0.1cm]
${\rm R}_{21}^{10}/R_{\rm L4}$   &  0.2$\pm$0.09  &  1.5$\pm$0.3   &  0.9$\pm$0.2   &  0.5$\pm$0.1   &  1.0           &  1.3$\pm$0.2   &  1.5$\pm$0.4   &  1.3$\pm$0.3  &   1.3$\pm$0.3  \\

\enddata 
 
\tablenotetext{(1-10)}{{\it First six rows:} Brightness temperature
  ratios between the different $^{12}$CO transitions derived for each
  line peak (Table~\ref{tab2}). We converted all $^{12}$CO integrated
  peak intensities from Jansky to Kelvin. {\it Rows seven to twelve:}
  Brightness temperature ratios with respect to peak L4,
  respectively.}

\label{tab4}
\end{deluxetable*}

\begin{deluxetable*}{lrrrrrrrrr}[!t]
\centering
\tablewidth{0pt}
\tablecaption{$^{12}$CO brightness temperature ratios derived from the sum of individual Gaussian Components. }
\tablehead{
\colhead{Ratio} & \colhead{L0} & \colhead{L1} & \colhead{L2} & \colhead{L3} 
& \colhead{L4} & \colhead{L5} & \colhead{L6} & \colhead{L7} & \colhead{L8} \\
     (1) & (2)            &  (3)	   & (4)    & (5)  & (6)  & (7)     
         & (8)            &  (9)           & (10)   }
\startdata 							  				

R$_{65}^{32}$                  & 1.50$\pm$0.40  & 2.00$\pm$0.30  & 1.80$\pm$0.20  & 1.90$\pm$0.30  & 1.40$\pm$0.10  & 1.80$\pm$0.20  & 2.00$\pm$0.30  & 3.30$\pm$0.60 &  9.60$\pm$3.90 \\[0.1cm]
R$_{65}^{21}$                  & 3.40$\pm$0.90  & 4.20$\pm$0.70  & 3.60$\pm$0.40  & 3.80$\pm$0.50  & 2.80$\pm$0.30  & 3.10$\pm$0.30  & 3.60$\pm$0.50  & 6.30$\pm$1.10 & 22.10$\pm$9.00 \\[0.1cm]
R$_{65}^{10}$                  & 0.60$\pm$0.20  & 2.10$\pm$0.40  & 1.70$\pm$0.20  & 1.70$\pm$0.20  & 1.50$\pm$0.20  & 1.60$\pm$0.20  & 2.20$\pm$0.30  & 3.40$\pm$0.60 & 10.70$\pm$4.40 \\[0.1cm]
R$_{32}^{21}$                  & 2.20$\pm$0.30  & 2.10$\pm$0.30  & 2.00$\pm$0.20  & 2.00$\pm$0.20  & 1.90$\pm$0.20  & 1.80$\pm$0.20  & 1.80$\pm$0.20  & 1.90$\pm$0.20 &  2.30$\pm$0.30 \\[0.1cm]
R$_{32}^{10}$                  & 0.40$\pm$0.10  & 1.10$\pm$0.10  & 0.90$\pm$0.10  & 0.90$\pm$0.10  & 1.10$\pm$0.10  & 0.90$\pm$0.10  & 1.10$\pm$0.10  & 1.00$\pm$0.10 &  1.10$\pm$0.20 \\[0.1cm]
R$_{21}^{10}$                  & 0.18$\pm$0.03  & 0.50$\pm$0.06  & 0.46$\pm$0.05  & 0.44$\pm$0.04  & 0.55$\pm$0.05  & 0.52$\pm$0.05  & 0.62$\pm$0.08  & 0.54$\pm$0.07 &  0.48$\pm$0.07 \\[0.3cm]
${\rm R}_{65}^{32}/R_{\rm L4}$   &  1.1$\pm$0.3   &  1.4$\pm$0.3   &  1.2$\pm$0.2   &  1.3$\pm$0.2   &  1.0           &  1.2$\pm$0.2   &  1.4$\pm$0.2   &  2.3$\pm$0.5  &   6.7$\pm$2.8  \\[0.1cm]
${\rm R}_{65}^{21}/R_{\rm L4}$   &  1.2$\pm$0.3   &  1.5$\pm$0.3   &  1.3$\pm$0.2   &  1.4$\pm$0.2   &  1.0           &  1.1$\pm$0.2   &  1.3$\pm$0.2   &  2.2$\pm$0.4  &   8.0$\pm$3.4  \\[0.1cm]
${\rm R}_{65}^{10}/R_{\rm L4}$   &  0.4$\pm$0.1   &  1.4$\pm$0.3   &  1.1$\pm$0.2   &  1.1$\pm$0.2   &  1.0           &  1.1$\pm$0.2   &  1.4$\pm$0.2   &  2.2$\pm$0.4  &   7.0$\pm$3.0  \\[0.1cm]
${\rm R}_{32}^{21}/R_{\rm L4}$   &  1.1$\pm$0.2   &  1.1$\pm$0.2   &  1.0$\pm$0.1   &  1.0$\pm$0.1   &  1.0           &  0.9$\pm$0.1   &  0.9$\pm$0.1   &  1.0$\pm$0.2  &   1.2$\pm$0.2  \\[0.1cm]
${\rm R}_{32}^{10}/R_{\rm L4}$   &  0.4$\pm$0.1   &  1.0$\pm$0.2   &  0.9$\pm$0.1   &  0.8$\pm$0.1   &  1.0           &  0.9$\pm$0.1   &  1.0$\pm$0.2   &  0.9$\pm$0.1  &   1.0$\pm$0.2  \\[0.1cm]
${\rm R}_{21}^{10}/R_{\rm L4}$   &  0.3$\pm$0.1   &  0.9$\pm$0.1   &  0.8$\pm$0.1   &  0.8$\pm$0.1   &  1.0           &  0.9$\pm$0.1   &  1.1$\pm$0.2   &  1.0$\pm$0.2  &   0.9$\pm$0.2  \\

\enddata

\tablenotetext{(1-10)}{{\it First six rows:} Brightness temperature
  ratios between the different $^{12}$CO transitions derived from the
  sum of all line peaks (Table~\ref{tab2}). We converted all $^{12}$CO
  integrated peak intensities from Jansky to Kelvin. {\it Rows seven
    to twelve:} Brightness temperature ratios with respect to peak L4,
  respectively.}

\label{tab4.5}
\end{deluxetable*}

Figure~\ref{fig8.0} shows the line spectra taken at the position of
each line peak for the four different $^{12}$CO transitions. Again,
the similar shape of the profiles amongst all four transitions for
each peak is striking. This is strong evidence that all four
transitions trace the same molecular gas, i.e., the same molecular
clumps along the disk of NGC~253. We fitted a multiple-component (up
to three) Gaussian to all line profiles. The results of these fits are
given in Table~\ref{tab3}. To reduce the parameter space for the fit
and to rest consistent between the fits for each transition, we fixed
the line centers of the three different line components by choosing
the best compromise for the line centers between the three lowest
transition after an initial fit to them with the line center as free
parameter. In Figure~\ref{fig8.0} it becomes clear why L6 appears
stronger in the velocity channel maps compared to L5 in the spectrally
integrated map: L5 is weaker but with a much broader line profile than
L6. Judging also from these complex line profiles with multiple
Gaussian components and given the high inclination of the disk in
NGC~253, it is very likely that each of these peaks is an ensemble of
GMCs at different velocities and merged along the line-of-sight. Based
on recent ALMA observations of multiple lines in the 3~mm band,
\cite{leroy15} find in total 10 different cloud components in their
data, as based on dense gas tracers such as HCN and HCO$^⁺$, which
appear to be fitted as well with up to three Gaussian components
similar to our work. Their main components along the disk are in good
agreement with those used in this study.

As a note, given the complexity of the peaks, we tried to run a clump
finder program on our data cubes \cite[e.g., {\tt GAUSSCLUMP}
from][]{stutzki1990}, but we did not manage to obtain meaningful
results due to the added complexity of an (almost) edge-on galaxy for
the dynamics as well as the lack of angular resolution.  Too many
clumps are still merged together in space {\it and} velocity so that a
clear separation and identification of GMCs remains very difficult at
this point.

\subsubsection{Line ratios}

\begin{figure*}[!t]
   \centering
   \resizebox{\hsize}{!}{\includegraphics{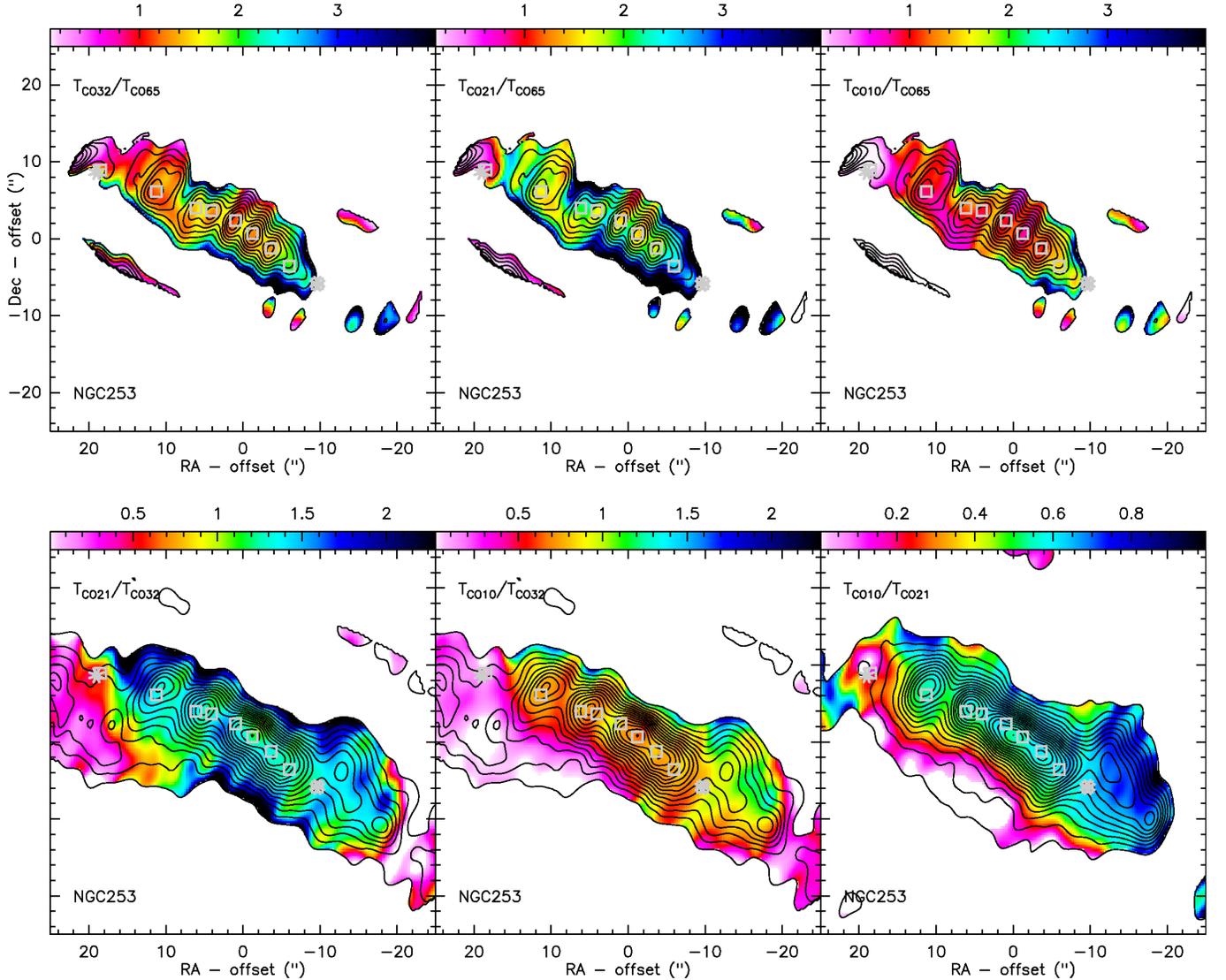}}
   \caption{Temperature brightness ratios between the different
     integrated $^{12}$CO line transitions. The grey symbols (squares
     and stars) are the same as in Fig.~\ref{fig3}. Contours in the
     upper panels are those of the \cosf\, emission while those in the
     lower panels are from the \cooz\, line emission. Please note that
     we used a reversed color scale, i.e., for low ratios bright
     colors such as white, pink and red are used while high ratios are
     represented with darker colors such as blue and black. This was
     done to facilitate the connection between warm (and/or dense)
     molecular gas that is reflected here in low line ratios while
     cooler (and/or less dense) gas is shown by large line ratios.}
   \label{fig9.0}
\end{figure*}

\begin{figure*}[!t]
   \centering
   \resizebox{\hsize}{!}{\includegraphics{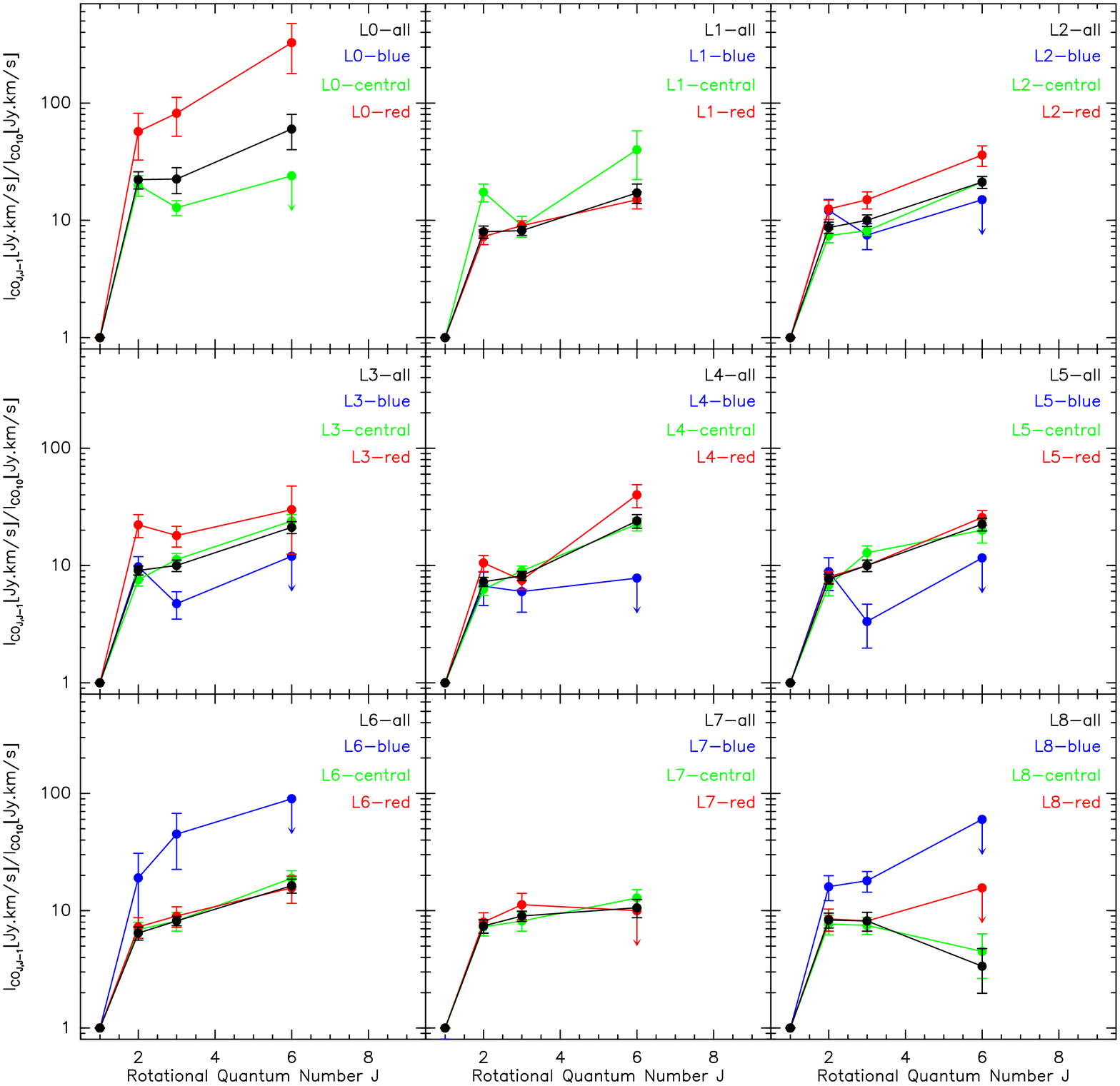}}
   \caption{CO line SEDs for each position (L0-L9) and Gaussian
     component (in black ({\it sum}), blue ({\it blueshifted}
     velocities), green ({\it central velocities}) and red ({\it
       redshifted} velocities)). The line intensities, from which the
     line ratios were derived and plotted here, were not converted
     into temperature scale (K.km/s) first, as opposed to
     Tables~\ref{tab4} and \ref{tab4.5}, in order to stay consistent
     with CO line SEDs from other publications
     \cite[e.g.,][]{papa10}.}
   \label{fig9.5}
\end{figure*}

\begin{figure*}[!t]
   \centering
   \resizebox{11.0cm}{!}{\includegraphics{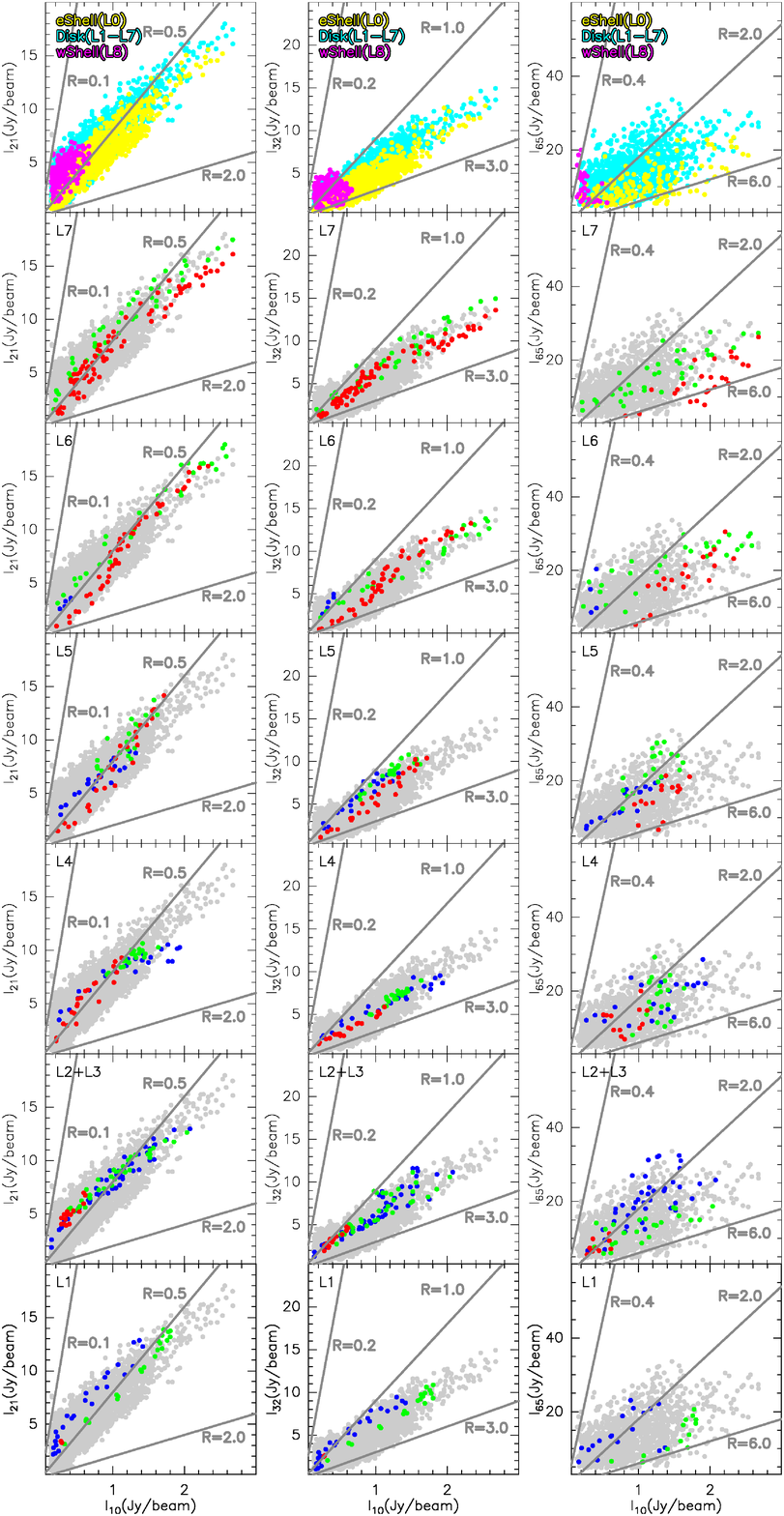}} 
   \caption{Comparison of $^{12}$CO line fluxes for the different line
     transitions ({\it left column: \coto\ vs. \cooz}, {\it middle
       column: \cott\ vs. \cooz}, {\it right column: \cosf\
       vs. \cooz}) and regions in NGC~253 (L1-L8; {\it from bottom to
       top}) derived from data cubes with matching spatial and
     spectral resolution (4$\farcs$2$\times$2$\farcs$1 and 12~\kms);
     we defined a region of $\sim$8$''$ (i.e. twice the synthesized
     beam) around each line peak defined in Fig.~\ref{fig3}. We used a
     sampling spacing of $\sim$4 pixels per beam. We only plot points
     that are above 3$\sigma$ for \cosf\, and 10$\sigma$ for the rest
     (due to the limited dynamical range).  The red-, blue- and
     green-colored points represent the blueshifted ($<-$40~\kms; in
     blue) and redshifted velocities ($>+$40~\kms; in red) and around
     the systemic velocity of NGC~253 (i.e., from $-$40~\kms\ to
     $+$40~\kms; in green), respectively. The grey data points are all
     data points, i.e. from all velocities and all regions. R
     represent the brightness temperature ratios and is defined as:
     R=T$_{\rm CO10}$/T$_{\rm COJ,J-1}$ for J$=$6,3,2. The solid lines
     are meant to guide the eyes for a given fixed ratio.}
   \label{fig13.0}
\end{figure*}

We calculated the brightness temperature ratios between \cosf, \cott,
\coto\, and \cooz, using different methods, first converting the
intensities from Jansky to Kelvin. Fig.~\ref{fig9.0} shows the ratios
between the spectrally integrated intensity maps from Fig.~\ref{fig3},
while Table~\ref{tab4} lists the ratios determined from the line
profiles shown in Fig.~\ref{fig8.0}. We plot the individual line SEDs
for each component and Gaussian component in Fig.~\ref{fig9.5}. We
also compare the line intensities sampled in bins of $\sim$1$''$ and
12~\kms\, in Fig.~\ref{fig13.0} by using a mask for the different
regions (being roughly twice the synthesized beam size); for
components L1 to L7 we plot the different velocities in different
colors to explore a possible dependence on velocity.

Along the disk, the spectrally integrated ratios (Fig.\ref{fig9.0})
show values of: R$_{32/65}^{\rm CO}$$\simeq$1-2, R$_{21/65}^{\rm
  CO}$$\simeq$2-3, R$_{10/65}^{\rm CO}$$\simeq$0.5-1.5,
R$_{21/32}^{\rm CO}$$\simeq$1-1.5, R$_{10/32}^{\rm CO}$$\simeq$0.5-1.5
and R$_{10/21}^{\rm CO}$$\simeq$0.3-0.7, the range of which is lower
than the values determined directly from the line components
(Table~\ref{tab4}). As Fig.~\ref{fig9.5} and Fig.~\ref{fig13.0}
clearly indicate, there are indeed differences between the three
different line components for each peak that appear to be identical in
the different line transitions. This highlights the fact that at each
position we probably see an ensemble of different GMCs which likely
exhibit slightly different excitation conditions (see also
Table~\ref{tab4.5}). However, the differences are not very large so
probably all of these GMCs are exposed to the same major heating
source, the starburst and its winds/shocks.

While we cannot identify any significant difference in the ratios
between the peaks along the disk (see also Columns 3-9 in
Table~\ref{tab4} \& \ref{tab4.5}, as well as Fig.~\ref{fig9.5}), the
temperature ratios are significantly lower close to the eastern shell
SB2 indicating much warmer (and/or denser) material there (similar to
the continuum emission) and slightly higher ratios toward the western
shell SB1 indicating less warm (or less dense) gas there (see
Figs.~\ref{fig9.0}, \ref{fig9.5} and \ref{fig13.0}).  This difference
in gas excitation between the two shells is quite surprising if one
assumes a similar nature/origin of the two. However, there are some
indications that the two shells are probably caused by different
mechanisms: SB2 is associated with winds from a stellar cluster while
SB1 is associated with a supernova remnant\footnote{Although one
  single supernova remnant is probably not enough to create SB1
  alone.}. The energetic output of a supernova is probably quite
different from the stellar winds from star-formation/star-bursts in a
stellar cluster so that a different feedback to the surrounding
molecular gas is to be expected (see also Section~3.2.4). Another
explanation could be that these shells are at different evolutionary
stages, although they appear to exhibit a similar extent and expansion
velocities \cite[see][]{saka06,bola13} if one believes that SB1 is an
expanding superbubble. As \cite{saka11} expressed doubts about SB1
being a superbubble as opposed to SB2, SB1 might have a much different
origin hence naturally explaining the different ratios and
subsequently different excitation conditions in these two regions.

However, we have to apply some caution in interpreting these ratios as
we do not have exactly the same uv-coverages in the different
$^{12}$CO transitions, with \cosf\, probably being the least well
sampled and the least sensitive. We already discussed the effect of
spatial filtering flux in a previous section and found that resolution
effects are possibly minor along the disk but could be more important
in the region of the shells. Therefore, we think that the ratios along
the disk are likely very representative while for the shells we
probably need more sensitive observations of a larger field with
better uv coverage, as is becoming possible with ALMA.

\begin{figure}[!t]
   \centering
   \resizebox{\hsize}{!}{\includegraphics{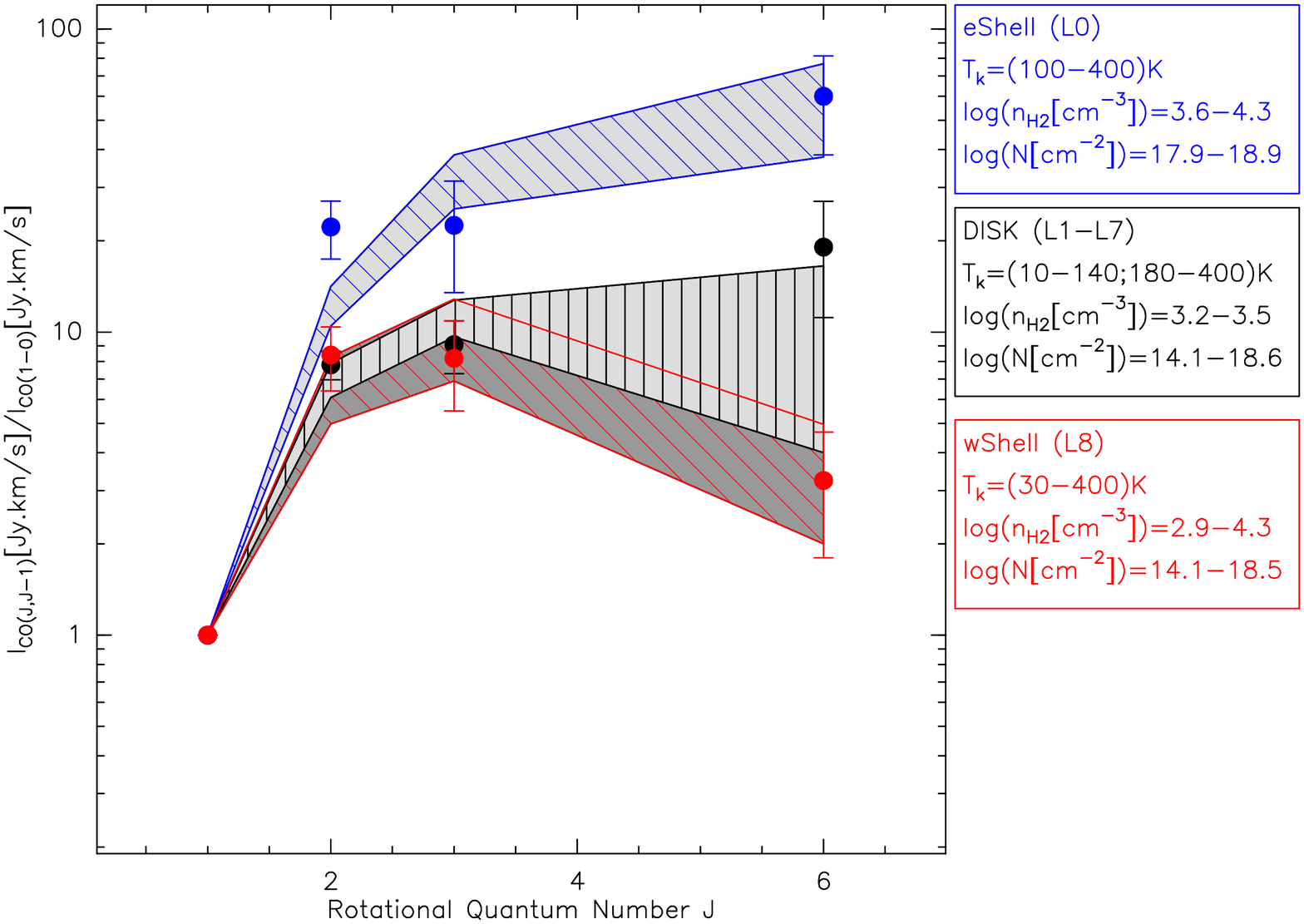}}
   \caption{CO line SEDs for the eastern ({\it blue filled circles})
     and western shell ({\it red filled circles}) and the central disk
     ({\it black filled circles}), summing up all Gaussian
     components. The {\it hashed grey areas} indicate the range of
     solutions with equally low $\chi_{\rm r}^2$-fits (i.e., solutions
     that are within $\sim$1-2$\sigma$ of the observed values) from
     RADEX simulations to the line ratios (here derived from the
     intensities in Jy.km/s as opposed to Table~\ref{tab4.5}, where
     they were derived from temperatures).  The best fit solutions
     intervalls are given in the small colored boxes to the right.}
   \label{fig9.75}
\end{figure}

\subsubsection{LVG Analysis}

\begin{figure}[!t]
   \centering
   \resizebox{6.89cm}{!}{\includegraphics{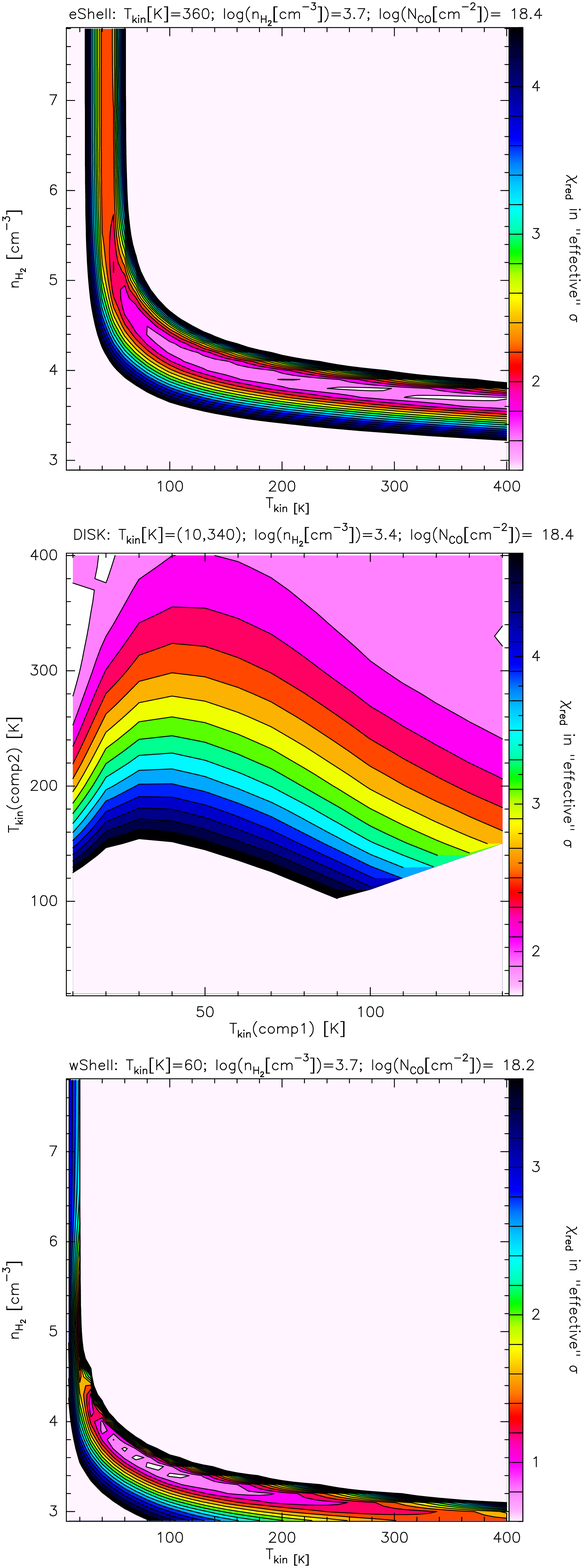}}
   \caption{Best fit solutions with the lowest $\chi_{\rm red}$ for
     the three different regions. For the one phase gas models of the
     western (wShell/L8) and eastern shells (eShell/L0; upper and
     lower panel) we plot the kinetic temperatures and CO column
     densities as function of $\chi^2_{\rm red}$ for a given H$_2$
     density while for the two phase gas model we show the two kinetic
     temperatures for a given H$_2$ density and CO column density
     (given as (T$_{\rm kin}$(comp1),T$_{\rm kin}$(comp2)) above the
     middle panel with T$_{\rm kin}$(comp1) for the ``cold'' and
     T$_{\rm kin}$(comp2) for the warm gas phase). The reduced
     $\chi_{\rm red}$ can be seen as effective $\sigma$s, meaning that
     a $\chi_{\rm red}$ of 1 (2,..)  represents modelled line ratios
     that are on average within 1$\sigma$ (2$\sigma$,..)  of the
     respective observed line ratios. The solution with absolute
     lowest $\chi_{\rm red}$ are given above each panel but note the
     range of possible solutions with equally low $\chi_{\rm red}$
     (these ranges are given in Fig.~\ref{fig9.75}). The regions in
     pale pink have $\chi_{\rm red}$ outside the range shown in the
     wedges and have been hence ``blanked''.}
   \label{fig9.85}
\end{figure}

We conducted an LVG analysis with the RADEX code \cite[see][]{vdt07}
applying a reduced $\chi^2$ minimization method to fit the basic
excitation conditions of the molecular gas in the western and eastern
shells and the (averaged) disk.  We first considered a one phase gas
model, i.e., gas at one kinetic temperature, one H$_2$ density and one
CO column density. We thereby varied the temperatures by T$_{\rm
  kin}=$10-400~K, the densities by n$_{\rm
  H_2}=$10$^3$-10$^8$~cm$^{-3}$ and the column densities by N$_{\rm
  CO}=$10$^{14}$-10$^{19}$~cm$^{-2}$ for a line width of dv=100~\kms\,
\cite[see also][]{leroy15} and a uniform sphere, linearly splitting up
the parameter space in 41 ``bins'' with steps of 10~K for T$_{\rm
  kin}$ and 0.1~dex for n$_{\rm H_2}$ and N$_{\rm CO}$. All reduced
$\chi^2_{\rm red}$ results for which the modelled line ratios are
within at least $\sim$1-1.5$\sigma$ of the observed line ratios (i.e.,
$\chi^2_{\rm red}\leq1.5$) were considered as acceptable
solutions. With these criteria we could fit the western and eastern
shell quite reasonably, while we did not find any solution for the
disk that could reproduce the high \cosf-to-\cooz\, ratio. This is in
agreement with previous findings on the CO line SED of NGC~253 by
different groups based on single-dish and/or satellite ({\it
  Herschel}) observations \cite[e.g.,][]{ros14,hail08}. One approach
is to allow for a two phase gas model by assuming two different
kinetic temperatures with a upper limit of the ``cold'' component at
around T$_{\rm kin, cold}$=150~K and T$_{\rm kin, warm}>$T$_{\rm kin,
  cold}$.  The range of solutions found for the excitation conditions
in the three different regions of NGC~253 are shown in
Fig.~\ref{fig9.75}. The respective range in kinetic temperatures for
the two component model (middle panel) for a given volume and column
density as well as the kinetic temperatures and column densities for
the one component models for a given density (upper and lower panel)
are plotted in Fig.~\ref{fig9.85}.

{\it Disk:} As mentioned just before, a one temperature phase model
largely underestimates the \cosf\, emission similar to previous
findings \cite[e.g.,][]{ros14,hail08}. Our best fit model suggests a
very cold gas component at around 10~K and a very warm one at around
340~K with H$_2$ gas densities for both of n$_{\rm
  H_2}$=10$^3$-10$^{4}$~cm$^{-3}$ and CO column densities of N$_{\rm
  CO}\simeq$10$^{18}$cm$^{-2}$ (see Fig.~\ref{fig9.85}).  However, as
can be seen in Fig.~\ref{fig9.75} and is indicated by the mediocre
$\chi^2$ values in Fig.~\ref{fig9.85}, we find a large range of
excitation conditions with similarly ``good'' $\chi^2$ values as the
best-fit model, certainly due to the fact that four observed lines are
not sufficient to constrain well the parameter space with four free
variables. Furthermore, although the best-fit models fit reasonably
well the observed data, the modeled line ratios are only within
1.5-2~$\sigma$ of the observed ones indicating that a two temperature
phase gas model is still insufficient to reproduce the data and the
assumption of equal gas and column densities of the two gas phases is
too simplistic. However, leaving those as free parameters as well
between the two gas phases enforces more observational constraints,
i.e., more line transitions and different molecular tracers need to be
observed which is out of the scope of this paper.

\cite{ros14} indeed argue that they need a three-temperature phase gas
model to explain their $^{12}$CO ladder, of which the phase-1 low
temperature component is heated through PDRs.  The phase-2 and 3
components are more likely heated mechanically while heating through
cosmic rays seems to be rather negligible based on their models
\cite[see next Section but also][]{pagl12}. The mechanical heating
through low-velocity shocks or turbulence is needed to produce the
high temperatures throughout the surrounding clouds. PDRs affect only
a thin layer of the surrounding molecular clouds and would potentially
start to dissociate the CO molecules due to the strong UV-radiation
fields \cite[see also:][]{ros14,goic13,kauf96}.

{\it Western Shell (L8):} Our best-fit solutions suggest for the
western shell two main scenarios due to the T$_{\rm kin}$-n$_{\rm
  H_2}$ degeneracy of $^{12}$CO in the excitation: either the
molecular gas is at 1.) rather high kinetic temperatures around 300~K
but low H$_2$ gas densities around $\sim$10$^3$~cm$^{-3}$, or 2.) at
lower kinetic temperatures around 60~K but high H$_2$ gas densities
around $\sim$5$\times$10$^3$~cm$^{-3}$ (see Fig.~\ref{fig9.75} and
Fig.~\ref{fig9.85}). Only in the low-T$_{\rm kin}$/high-n(H$_2$) case,
the CO column density can be well constrained and show values of about
$\sim$10$^{¹8}$~cm$^{-2}$, similar to the disk and the eastern shell
(see Fig.~\ref{fig9.85}), while the high-T$_{\rm kin}$/low-n(H$_2$)
case leaves open almost the entire range of column densities sampled
here. The HCN emission has been found to be stronger (by a factor of
$\sim$3 in integrated fluxes) in the western shell compared to the
eastern shell by \cite{leroy15}.  This is likely incompatible with the
case in which the gas in the western shell is significantly less dense
than in the eastern shell, favoring hence the low-T$_{\rm
  kin}$/high-n(H$_2$) case. The kinetic temperature in this case of
60~K is in between the cold (T$_{\rm kin}\simeq$10~K) and hot (T$_{\rm
  kin}\simeq$340~K) gas component found for the disk but much lower
than found for the eastern shell (see next paragraph).

  {\it Eastern Shell (L0):} The best fit solution for the eastern
  shell suggests higher kinetic temperatures (T$_{\rm kin}$=360~K)
  than the disk and the western shell but at similar H$_2$ densities
  (n$_{\rm H_2}\simeq5\times10^3$~cm$^{-3}$) and CO column densities
  (N$_{\rm CO}\simeq$10$^{18}$~cm$^{-2}$). However, it seems that the
  best fit model slightly underestimates the \coto\, emission as
  indicated by the lowest $\chi^2_{\rm red}$ of $\sim$1.2 which means
  that not all modelled line ratios lie within 1$\sigma$ of the
  observed ones. We also ran a two-temperature phase gas model for the
  eastern shell similar to the disk but we did not find a solution in
  which the \coto\, emission was correctly reproduced. The best fit
  solutions of the two-temperature phase models largely resembled the
  one-temperature phase models so that we hence decided to keep the
  one-temperature phase model for the eastern shell for simplicity
  reasons. Either we are seeing (part of) the warm component from the
  disk only here or additional excitation mechanisms have to be
  considered.

\subsubsection{Comparison to other multi-transition
  CO studies}

Studies of spatially resolved, multi-transition ($\geq4$) $^{12}$CO
emission in galaxies are still very rare. Most of them are conducted
with single dish telescopes that provide sufficient angular resolution
to resolve GMC scales ($\leq$50-100pc) only in the most nearby
galaxies or our own galaxy, if at all. We picked several $^{12}$CO
ladder studies in order to compare them to the results obtained on
NGC~253
\cite[][]{topa14,meij13,goic13,dani11,vdwer10,riech10,cari10,empr09}. The
studied objects span from GMCs in the galactic center, over nearby
starburst and/or AGN galaxies such as NGC6946, M82 and NGC~1068 to
ULIRGs like Arp~220 and Mrk~231 and
high-redshift galaxies.\\
\\
{\it Disk:} Interestingly, all studies, i.e., for all (active)
environments in our vicinity up to the high-redshift universe, seem to
need a multi-phase gas model (with at least a high and lower
temperature gas component) to be able to reproduce the observed
multi-transition CO emission
\cite[e.g.,][]{topa14,goic13,dani11,vdwer10,riech10,cari10,empr09,weis05},
similar to the central disk in NGC~253.  Also, most of them find that
besides photoelectric heating from UV radiation in PDRs also
mechanical heating through low-velocity shocks have to be considered
as almost equally dominant mechanism to heat the molecular gas
\cite[see also ][]{garc14,krip11,garc10}.  In all these regions,
cosmic rays do not seem to contribute significantly to the heating of
the warm molecular gas component, agreeing with the findings by
\cite{ros14} \cite[see also][]{meij11}.  On the other hand, NGC~253
has known supernova explosions at a rate of $<$0.2~yr$^{-1}$
\cite[see][]{rampa14,pagl12} and, more importantly, a median cosmic
ray ionisation rate of $\zeta_{\rm CR}\simeq$6$\times10^{-12}{\rm
  s}^{-1}$ \cite[][]{pagl12}. This is a factor of $\sim$100-1000
higher than found in the vicinity of Sgr~A$^\star$ \cite[][]{goto08}
and of the order of the X-ray(/cosmic ray) ionisation rates of
$>$10$^{-13}$s$^{-1}$ found in the (U)LIRGs Arp~220 and NGC~4418
\cite[][]{gonz13}. \cite{goic13} mention that already a cosmic ray
ionisation rate of $\zeta_{\rm CR}\simeq10^{-14}{\rm s}^{-1}$ can heat
the gas up to $\gg$10~K and \cite{pagl12} conclude that ``cosmic ray
penetration and heating is an important contributor to warm
temperatures observed in starburst galaxies''. Although the
simulations conducted by \cite{ros14} favor a minor role of cosmic ray
heating, it probably cannot be completely excluded at this
point. Moreover, in the case of the ULIRG Mrk~231, \cite{vdwer10} find
indications that, besides PDR heating, X-ray(/cosmic ray) heating is a
significant contributor in order to explain their two-temperature
phase model of the multi-transition CO emission although they do not
include shock models in their simulations. On the other hand,
\cite{rang11} find in the case of Arp~220 that mechanical heating
should be largely dominating over that by PDR, XDR and cosmic rays, at
least for the warm molecular gas. This result might be logical as
Arp~220 is a known major merger in an evolved merging stage so that
mechanical heating should be quite significant. Although in a slightly
earlier merger stage than Arp~220, \cite{meij13} argue for a (low
velocity, C-type) shock domination of the gas heating in NGC~6240 as
well, similar to Arp~220.

In contrast to NGC~1068, whose molecular gas in its central
$\sim$200pc disk is most likely dominated by a combination of (low
velocity) shocks and a giant XDR \cite[e.g.,][]{krip11,garc14,viti14},
the line transition ratios in NGC~253 decrease (assuming CO(J$_{\rm
  upper}$-J$_{\rm lower}$)/CO(J=1$-$0); with J$_{\rm upper}>1$) while
those for NGC~1068 increase. This is in good agreement with
theoretical predictions of the $^{12}$CO line ratios for XDRs and PDRs
\cite[][]{meij06}, even in the case of a strong cosmic ray ionisation
field in addition to a PDR. Another interesting difference to note is
that NGC~1068 indeed shows variations of its line ratios the closer
one gets to the nucleus (central 50~pc or so) as revealed by recent
ALMA Observations \cite[see][]{garc14,viti14} while the disk (central
500~pc) appears to be dominated by the same heating source. Hence, a
starburst influences its surrounding gas on a much larger scale than
an AGN seems to do. \\
\\
{\it Shells:} Based on the discussion of the heating mechanisms
responsible for the CO ladder in diverse environments, one can draw
some first conclusions on the dominant heating mechanisms for the two
shells by simple comparison. As mentioned before, probably only
mechanical energy from shocks or turbulent gas can heat up the bulk of
molecular gas significantly above 50~K, while photo-electric (PDR)
and/or cosmic ray heating dominates the cold to warm gas
component. Our data quite clearly indicate that the eastern shell is
much warmer than the disk and the western shell while being at similar
gas and column densities.  This could be hence an indication that
shocks might play a much more dominant role in the eastern shell,
associated with a compact stellar cluster, than in the PDR+shock
dominated disk and the western shell which might be connected to a
supernova remnant. It appears rather unlikely that a strong PDR (or
cosmic rays) also has to be considered for the eastern shell which
could explain the stronger emission in the higher-J transitions in it
compared to the disk; the higher-J CO transitions are more dominantly
excited than the lower-J CO transitions.

Our data suggest that the western shell is much colder than the
eastern shell and the hot gas component of the disk but is still
warmer than the cold gas component of the disk.  One conclusion,
although certainly quite speculative at this point, is that the cosmic
rays from the supernova remnant dominate the gas excitation in the
western shell and that the (low-velocity) shocks, if they exist, from
the SN explosion have not yet or only very inefficiently heated the
gas to the same high temperatures as found in the (warm component of
the) disk or the eastern shell. However, given the similar extent of
both shells, it is not intuitive why shocks play such a minor role in
the western shell compared to the eastern shell and the disk.  Also,
while it is true that SN have been identified through radio
observations in the western shell, the lack of radio detections in the
eastern shell does not exclude SN in it. Further observations of
either $^{12}$CO or other shock and cosmic ray tracers have to be
conducted to unveil the heating mechanisms in these shells at a
sufficient confidence level.


\section{Summary}
\label{sum}

We presented new interferometric observations of the extended \cosf\,
line and 686~GHz continuum emission in NGC~253 carried out as a five
point mosaic with the SMA. These data were then compared to three
lower-J $^{12}$CO transitions (J=1--0, J=2--1, J=3--2) and continuum
emission at lower frequencies at similar angular resolution of $\sim$
4$''$ observed with ALMA and the SMA. 

Five of the eight $^{12}$CO line peaks find counterparts in the mm and
submm continuum emission underlying the thermal nature of the
continuum emission from dust as suggested by the spectral index of 3
determined from the different frequencies. The continuum emission at
the lowest frequency of 115~GHz appears to exhibit already a
significant contribution from non-thermal synchrotron and thermal
free-free processes (between 15-80\%) as supported by the spectral
indices derived at cm wavelengths. We derive dust temperatures of
$\sim$10-25~K for the different continuum peaks in the disk with low
opacities of $\sim$0.2 at 686~GHz. However, we also find indications
for a hotter dust component at least in the inner disk of NGC~253 with
dust temperatures exceeding 60~K and at much larger opacities of
around 4 at 686~GHz. The latter dust component approaches the values
found for ULIRGs such as Arp~220. We estimate a total dust mass of a
few 10$^6$~M$_\odot$ for the entire disk, splitting up into
$\sim$10$^4$-10$^5\odot$ for the GMCs at the individual continuum
peaks. Based on the gas masses of each of the peaks, we find a
gas-to-dust mass ratio of the order of $\sim$100-1000.

The \cosf\, emission follows nicely the distribution of the molecular
gas seen in the lower-J transitions with roughly eight peaks along the
disk. Only few \cosf\, emission is detected close to the two shells
emerging from the edges of the central disk. While the $^{12}$CO line
transition ratios do not vary significantly along the disk, the two
shells show quite different ratios not only compared to the disk but
also to each other. The line ratios found along the disk seem to
necessitate a two-phase gas model in agreement with previous studies
on NGC~253 as well as multiple-$^{12}$CO observations on other active
galaxies. This two-phase gas model is mainly based on two
temperatures, a cool gas component at around T$_{\rm kin}$=10~K and
hot gas component at around T$_{\rm kin}$=300~K.  Following a similar
argumentation to previous publications, the lower temperature gas is
probably dominated by the PDR while the higher temperature gas is a
consequence of the shocks found throughout the disk \cite[see
also][]{ros12,ros14}.  However, a possible contribution from cosmic
rays, given the high cosmic ray ionisation rate within the disk,
cannot be completely excluded for either temperature phase at this
point although some simulations indicate only a minor role of cosmic
ray heating \cite[e.g.,][]{ros14}. While the eastern shell exhibits
even warmer gas (T$_{\rm kin}>$300~K) with respect to the hot gas
(T$_{\rm kin}\simeq$300~K) component of the disk, the western shell
contains gas much cooler (T$_{\rm kin}\simeq$60~K) than the eastern
shell but somewhere in between the two temperature gas components of
the disk (T$_{\rm kin}^{cold}\simeq$10~K and T$_{\rm
  kin}^{hot}\simeq$300~K); the gas densities
(n(H$_2$)$\simeq$5$\times$10$^{3}$~cm$^{-3}$) and column densities
(N(CO)$\simeq$10$^{18}$~cm$^{-2}$) are very similar between the two
shells and the disk. This reflects either a different evolutionary
stage of the shells, an additional, different or more efficient
heating mechanism in the eastern shell, a very different nature of
these two structures, or a combination thereof. However, follow-up
observations are mandatory to put our findings for the shells onto a
more solid basis.

\acknowledgments

The Submillimeter Array is a joint project between the Smithsonian
Astrophysical Observatory and the Academia Sinica Institute of
Astronomy and Astrophysics and is funded by the Smithsonian
Institution and the Academia Sinica.  This paper makes use of the
following ALMA data: ADS/JAO.ALMA\#2011.0.00172.S. ALMA is a
partnership of ESO (representing its member states), NSF (USA) and
NINS (Japan), together with NRC (Canada) and NSC and ASIAA (Taiwan),
in cooperation with the Republic of Chile. The Joint ALMA Observatory
is operated by ESO, AUI/NRAO and NAOJ. K.S. was supported by the grant
NSC 102-2119-M-001-011-MY3. The National Radio Astronomy Observatory
is a facility of the National Science Foundation operated under
cooperative agreement by Associated Universities, Inc.  We thank the
anonymous referee for the thourough and constructive comments.

{\it Facilities:} \facility{SMA}.\facility{ALMA}.

\end{document}